\documentclass[12pt]{article}

\usepackage[dvips]{graphicx}
\usepackage{amsmath}
\usepackage{amssymb}

\DeclareSymbolFont{AMSb}{U}{msb}{m}{n}
\DeclareMathSymbol{\N}{\mathbin}{AMSb}{"4E}
\DeclareMathSymbol{\Z}{\mathbin}{AMSb}{"5A}
\DeclareMathSymbol{\R}{\mathbin}{AMSb}{"52}
\DeclareMathSymbol{\Q}{\mathbin}{AMSb}{"51}
\DeclareMathSymbol{\I}{\mathbin}{AMSb}{"49}
\DeclareMathSymbol{\C}{\mathbin}{AMSb}{"43}

\begin{document}

\begin{center}

\vspace{0.5cm}
\textbf{\Large Fast computation of the Gauss hypergeometric function with all its parameters complex with
application to the P{\"o}schl-Teller-Ginocchio potential wave functions}

\vspace{5mm} {\large N.~Michel \footnote{\textit{E-mail address:} nmichel@yukawa.kyoto-u.ac.jp \\
\hspace*{0.5cm} \textit{Phone : 81-75-753-3857} \\
\hspace*{0.5cm}
\textit{Fax : 81-75-753-3886}}$^{,a}$ and M.V.~Stoitsov$^{~a, b}$}

\vspace{3mm}
{$^a$ \textit{Department of Physics, Graduate School of Science, \\ Kyoto University, Kitashirakawa, Kyoto, 606-8502, (Japan)} \\
 $^b$ \textit{Department of Physics \& Astronomy, \\
 University of Tennessee, Knoxville, Tennessee 37996, USA  \\
 Physics Division, Oak Ridge National Laboratory, \\
 P.O.~Box 2008, Oak Ridge, Tennessee 37831, USA \\
 Joint Institute for Heavy-Ion Research, Oak Ridge, Tennessee 37831, USA \\
 Institute of Nuclear Research and Nuclear Energy, \\
 Bulgarian Academy of Sciences, Sofia-1784, Bulgaria}
}
\end{center}

\vspace{5mm}

\hrule \vspace{5mm} \noindent{\Large \bf Abstract}
\vspace*{5mm}

The fast computation of the Gauss hypergeometric function $_2F_1$ with all its parameters complex is a difficult task. 
Although the $_2F_1$ function verifies numerous analytical properties involving power series expansions whose implementation is apparently immediate, 
their use is thwarted by instabilities induced by cancellations between very large terms. 
Furthermore, small areas of the complex plane, in the vicinity of $\displaystyle z = e^{\pm i \frac{\pi}{3}}$,
are inaccessible using $_2F_1$ power series linear transformations. In order to solve these problems, 
a generalization of R.C.~Forrey's transformation theory has been developed. 
The latter has been successful in treating the $_2F_1$ function with real parameters. As in real case transformation theory, 
the large canceling terms occurring in $_2F_1$ analytical formulas are rigorously dealt with, 
but by way of a new method, directly applicable to the complex plane. 
Taylor series expansions are employed to enter complex areas outside the domain of validity of power series analytical formulas. 
The proposed algorithm, however, becomes unstable in general when $|a|,|b|,|c|$ are moderate or large. 
As a physical application, 
the calculation of the wave functions of the analytical P{\"o}schl-Teller-Ginocchio potential involving $_2F_1$ evaluations is considered.

\vspace{5mm} \noindent{\Large \bf Program Summary}

\bigskip\noindent{\it Title of the programs:} hyp\_2F1, PTG\_wf

\bigskip\noindent{\it Catalogue number:}

\bigskip\noindent{\it Program obtainable from:}
CPC Program Library, \
Queen's University of Belfast, N. Ireland

\bigskip\noindent{\it Program summary URL:}

\bigskip\noindent{\it Licensing provisions:} none

\bigskip\noindent{\it Computers on which the program has been tested:} Intel i686

\bigskip\noindent{\it Operating systems:} Linux, Windows

\bigskip\noindent{\it Programming language used:} C++, Fortran 90

\bigskip\noindent{\it Memory required to execute with typical data:}

\bigskip\noindent{\it No.\ of bits in a word:} 64

\bigskip\noindent{\it No.\ of processors used:} 1

\bigskip\noindent{\it Has the code been vectorized?:} No

\bigskip\noindent{\it No.\ of bytes in distributed program, including test data, etc.:}

\bigskip\noindent{\it No.\ of lines in distributed program:}

\bigskip\noindent{\it Nature of physical problem:}
The Gauss hypergeometric function $_2F_1$, with all its parameters complex, is  
uniquely calculated in the frame of transformation theory with power series summations, 
thus providing a very fast algorithm. The evaluation of the wave functions of the analytical 
P{\"o}schl-Teller-Ginocchio potential is treated as a physical application.

\bigskip\noindent{\it Method of solution:}
The Gauss hypergeometric function $_2F_1$ verifies linear transformation formulas allowing to consider arguments of a small modulus 
which then can be handled by a power series. They, however, give rise to indeterminate or numerically unstable cases, 
when $b-a$ and $c-a-b$ are equal or close to integers. They are properly dealt with through analytical manipulations of the 
Lanczos expression providing the Gamma function. The remaining zones of the complex plane uncovered by transformation formulas 
are dealt with Taylor expansions of the $_2F_1$ function around complex points where linear transformations can be employed. 
The P{\"o}schl-Teller-Ginocchio potential wave functions are calculated directly with $_2F_1$ evaluations.

\bigskip\noindent{\it Restrictions on the complexity of the problem:}
The algorithm provides full numerical precision in almost all cases for $|a|$, $|b|$, and $|c|$ of the order of one or smaller, 
but starts to be less precise or unstable when they increase, especially through $a$, $b$, and $c$ imaginary parts. 
While it is possible to run the code for moderate or large $|a|$, $|b|$, and $|c|$ and obtain satisfactory results for some specified values, 
the code is very likely to be unstable in this regime.

\bigskip\noindent{\it Typical running time:} 20,000 $_2F_1$ function evaluations take an average of one second.

\bigskip\noindent{\it Unusual features of the program:} Two different codes, one for the hypergeometric function
and one for the P{\"o}schl-Teller-Ginocchio potential wave functions, are provided in C++ and Fortran 90 versions.



\bigskip

\vspace{5mm}
\noindent{\Large \bf Long Write-up}
\vspace*{5mm}

\section{Introduction}

The Gauss hypergeometric function $_2F_1$ is one of the most important special functions in complex analysis. 
Due to its three defining parameters, its structure is extremely rich and contains almost all special functions as a particular or limiting case. 
Important orthogonal polynomials, i.e.~Chebyshev, Legendre, Gegenbauer, and Jacobi polynomials, can all be expressed with the $_2F_1$ function. 
It is also the case for elementary functions such as power functions, $\log$, $\arcsin$ and $\arctan$, 
and more complicated functions such as elliptic integrals. Letting one of its parameters go to $+\infty$, one can obtain,  
as a limiting case, the confluent hypergeometric function of the first and second kind, which itself contains, 
as particular cases, exponential and sine/cosine elementary functions, as well as Bessel, Hankel, and Coulomb functions. 
(See Ref.\cite{cwfcomplex} for a recent implementation by the first author of the confluent hypergeometric function in the context of Coulomb wave functions.)

Its physical applications are plethora. For example, it is well known that integrals in atomic collision descriptions 
are very often analytically expressed with the $_2F_1$ function \cite{Alder_Bohr,coulex_book,coulex_paper,van_ingen,Datta}. 
It also appears in the Kepler problem, classically perturbed \cite{Kepler_classical} and relativistic \cite{Kepler_relativistic}. 
Still in the domain of astrophysics, it can serve as a solution in linearized magnetohydrodynamics equations \cite{linear_MHD}, 
and is found in the study of gravitational waves emission by binary stars, where its computation has to be very precise \cite{bin_stars_grav_waves}. 
The $_2F_1$ function plays a particularly important role in the topic of analytical potentials. 
Indeed, wave equations seldom admit analytical solutions, 
so that the study of potentials giving rise to wave functions expressible in closed form is very much involved. 
It can provide analytical wave function solutions of the Schr{\"o}dinger equation 
\cite{Eckart_pot, Rosen_Morse_pot, Hulthen_pot, Manning_Rosen_pot1, Manning_Rosen_pot2, Natanzon_pot1} 
and of the relativistic Klein-Gordon \cite{Klein-Gordon} and Dirac equations \cite{Dirac1,Dirac2,Dirac3}. 
At the non-relativistic level, the hypergeometric potentials of Natanzon type are particularly important due to their six defining parameters, 
making them very flexible \cite{Natanzon_pot1,Ginocchio_SUSY,Natanzon_SO21,Natanzon_path_int}. 
They contain the P{\"o}schl-Teller-Ginocchio (PTG) potential as a particular case \cite{PTG_pot,PTG_path_int}, 
which is widely used in nuclear physics due to its flat shape mimicking the nuclear interior \cite{Karim,Mario1,Mario2,Mario3}. 
Furthermore, it bears bound, resonant, or scattering states, features very interesting for the study of loosely bound or resonant nuclei. 
As its wave functions contain $_2F_1$ function components and also because their computation is not trivial, 
a section of this paper and part of the published code are dedicated to this problem as a direct application.

The numerical computation of the $_2F_1$ function is a challenging task. 
Firstly, the power series defining $_2F_1$ has a radius of convergence equal to one, 
which implies it can be used only for $z$ arguments of small modulus. 
While transformations formulas can minimize the modulus of the argument \cite{Abramowitz_2F1}, 
they cannot be tackled directly in numerical applications because of the appearance of canceling diverging terms 
for $b-a$ and $c-a-b$ in the vicinity of integers \cite{Forrey}. 
Moreover, they cannot span all the complex plane, because the points $\displaystyle z = e^{\pm i \frac{\pi}{3}}$ cannot be reached with any of them \cite{Temme_pres,Buhring}.
Added to that, the case of even moderate $|a|,|b|$, and $|c|$ is problematic due to the rapid growth of power series terms with $a, b$, and $c$ \cite{Temme_pres}. 
Analytic continuation of the $_2F_1$ function by use of convergent series in the complex plane has been treated in Ref.\cite{Becken_Schmelcher},
where all possible transformations are devised. However, aforementioned problems remain in their approach, as numerical problems occur not for integer $b-a$ and $c-a-b$,
for which analytic solutions were presented, but for $b-a$ and $c-a-b$ in the vicinity of integers.
For the case of real arguments, 
R.C.~Forrey's transformation theory \cite{Forrey} has allowed to remove the divergences induced by near-integer $b-a$ and $c-a-b$ at the price, 
however, of very complicated formulas. They rely on Chebyshev polynomial expansions of the Gamma function for arguments belonging to $[-\frac{1}{2}:\frac{3}{2}]$, 
real numbers outside this interval being handled through the standard Gamma function property $\Gamma(x+1) = x \Gamma(x)$. 
This algorithm is, however, clearly impossible to use in the complex plane. 
The only code calculating the $_2F_1$ function in the complex plane known by the authors is the one of Numerical Recipes \cite{NR}, 
which integrates numerically the hypergeometric equation in the complex plane starting from a $_2F_1$ value of its disk of convergence. 
While all problems mentioned above are removed with this method, it is necessarily slower in practice than direct summation of power series and, moreover, 
demands a careful check of the used complex path of integration, 
as integrated functions should never decrease rapidly on their path to avoid instabilities \cite{NR}. 
Added to that, full numerical precision can never be reached with direct integration \cite{NR}. 
In order to have both fast and accurate computation of the $_2F_1$ function in the complex plane, the authors chose to use R.C.~Forrey's transformation theory, 
but modified so it relies on the mathematical structure of the Lanczos numerical approximation of the Gamma function. 
The Chebyshev method problem is then avoided, and formulas are much simpler as well.
The missing parts of the complex plane, including the two important points $\displaystyle z = e^{\pm i \frac{\pi}{3}}$, are precisely handled by Taylor expansions of the $_2F_1$ function, 
so that the evaluation of $_2F_1$ in the remaining zones of the complex plane costs only two normal $_2F_1$ function calls 
and one quickly converging Taylor series evaluation. 
They can also be handled with non-hypergeometric power series expansions \cite{Buhring}, but they are available for non-integer or vanishing $b-a$ only.
While this method is very efficient for $|a|,|b|$, and $|c|$ of the order of one or smaller, 
the general case of moderate or large $|a|,|b|$, and $|c|$ with $z$ complex remains, however, problematic.

\section{Definition of the Gauss hypergeometric function}
One defines the Gauss hypergeometric function $_2F_1$ through its power series expansion around $z=0$ \cite{Abramowitz_2F1}:

\begin{eqnarray}
_2F_1(a,b,c;z) &=& \sum_{n=0}^{+\infty} \frac{(a)_n (b)_n}{(c)_n n!} z^n \label{hyp_PS_zero} \\
&=& \sum_{n=0}^{+\infty} t_n z^n  \label{hyp_PS_zero_2} \\
t_{0} &=& 1 \nonumber \\
t_{n+1} &=& \frac{(a+n)(b+n)}{(c+n)(n+1)} t_n \mbox{ , } n \geq 0 \label{hyp_PS_zero_term}
\end{eqnarray}
where $\displaystyle (x)_n = 1 \cdot x \cdots (x+n-1)$ is the standard Pochhammer symbol.
The recursive definition of the series term, used in numerical calculations, is also provided in Eqs.(\ref{hyp_PS_zero_2},\ref{hyp_PS_zero_term}).
$a$ and $b$ can be arbitrary complex numbers and the series reduces to a polynomial if $a \in \Z^{-}$ or $b \in \Z^{-}$.
The domain of definition of $c$ depends on $a$ and $b$: one can have either $\displaystyle c \in \C \backslash \Z^{-}$, or $c \in \Z^{-}$ as long as
$a \in \Z^{-}$ or $b \in \Z^{-}$, so that $c < a$ or $c < b$.
In this last case, the power series terminates before $(c)_n$ can be equal to zero.

This series always converges for $|z| < 1$, conditionally for $|z| = 1$, and always diverges for $|z| > 1$.
${_2F_1}(a,b,c;z)$ is defined in the rest of the complex plane by analytic continuation, where it is defined everywhere
except on its cut on $]1:+\infty[$. From now on, we will denote ${_2F_1}(a,b,c;z)$ with its other standard notation $F(z)$, unless necessary.
$F(z)$ verifies the following differential equation:

\begin{eqnarray}
&&z(1-z)F''(z) + [c-(a+b+1)z] F'(z) - ab F(z) = 0 . \label{hyp_eq}
\end{eqnarray}

In order to have a solution of Eq.(\ref{hyp_eq}) in the case of undefined $F(z)$, thus with $c \in \Z^{-}$,
one uses the limit of the function $\displaystyle \frac{F(z)}{\Gamma(c)}$ for $c \rightarrow -m$, with $m \in \N$
and $\Gamma(z)$ the standard Gamma function \cite{Abramowitz_2F1,Abramowitz_Gamma}:

\begin{eqnarray}
&&\lim_{c \rightarrow -m} \frac{F(z)}{\Gamma(c)} = \frac{(a)_{m+1} (b)_{m+1}}{(m+1)!} z^{m+1} {{_2F_1}}(a+m+1,b+m+1,m+2;z) .
\end{eqnarray}

Eq.(\ref{hyp_eq}) provides a useful test $T$ of the accuracy of the numerical evaluation of $_2F_1$ functions for $z \not\in \{0,1\}$:
\begin{eqnarray}
&&T = \frac{\left| F''(z) + \frac{\displaystyle [c-(a+b+1)z] F'(z) - ab F(z)}
{\displaystyle z(1-z)} \right|_{\infty}}{|F(z)|_{\infty} + |F'(z)|_{\infty} + |F''(z)|_{\infty}} \label{2F1_test} .
\end{eqnarray}

$F(0)$ is trivially tested as $F(0) = 1$, whereas the following $T_1$ test holds for $F(1)$ as long as $F(1)$ and $F'(1)$ are defined:
\begin{eqnarray}
&&T_1 = \frac{\left| \displaystyle [c-(a+b+1)] F'(1) - ab F(1) \right|_{\infty}}
{\displaystyle |F(1)|_{\infty} + |F'(1)|_{\infty} + 10^{-307}} \label{2F1_test_one} .
\end{eqnarray}

\section{Basic linear transformation formulas} \label{linear_transf_section}
In order to be able to calculate $F(z)$ for $|z| \geq 1$, one will follow Ref.\cite{Forrey} by employing
linear transformation formulas. They have the peculiarity to express $F(z)$ with ${_2F_1}$ functions with
other parameters and arguments $a$, $b$, $c$, and $z$, so that it is almost always possible to reach arguments of modulus small enough
so that the ${_2F_1}$ functions issued from the transformation can be evaluated with a power series.
The linear transformation formulas used in the code consist in \cite{Abramowitz_2F1}:

\begin{eqnarray}
{_2F_1}(a,b,c;z) &=& {_2F_1}(b,a,c;z)  \label{T_ba} \\
&=& (1-z)^{c-a-b} {_2F_1}(c-a,c-b,c;z) \label{T_cab} \\
&=& (1-z)^{-a} {_2F_1} \left( a,c-b,c;\frac{z}{z-1} \right) \label{T_ma} \\
&=& (1-z)^{-b} {_2F_1} \left( b,c-a,c;\frac{z}{z-1} \right) \label{T_mb} \\
&=& \frac{\Gamma(c) \Gamma(c-a-b)}{\Gamma(c-a) \Gamma(c-b)} {_2F_1}(a,b,a+b-c+1;1-z) \nonumber \\
&+& (1-z)^{c-a-b} \frac{\Gamma(c) \Gamma(a+b-c)}{\Gamma(a) \Gamma(b)} {_2F_1}(c-a,c-b,c-a-b+1;1-z) \label{T_mzp1} \\
&=& \frac{\Gamma(c) \Gamma(b-a)}{\Gamma(b) \Gamma(c-a)} (-z)^{-a} {_2F_1} \left( a,1-c+a,1-b+a;\frac{1}{z} \right) \nonumber \\
&+& \frac{\Gamma(c) \Gamma(a-b)}{\Gamma(a) \Gamma(c-b)} (-z)^{-b} {_2F_1} \left( b,1-c+b,1-a+b;\frac{1}{z} \right) . \label{T_z_inv}
\end{eqnarray}

The first four transformations of Eqs.(\ref{T_ba},\ref{T_cab},\ref{T_ma},\ref{T_mb}) are always numerically stable.
Problems, however, arise with the transformations of Eqs.(\ref{T_mzp1},\ref{T_z_inv}). Indeed, $\Gamma(z)$ is undefined for negative integer arguments,
so that these two expressions are indeterminate if $c-a-b$ for Eq.(\ref{T_mzp1}) or $b-a$ for Eq.(\ref{T_z_inv}) are integers.
The case of negative integer $c$ is not important, as these formulas are not used in practice for $F(z)$ being polynomial. As ${_2F_1}(a,b,c;z)$ can be perfectly
defined even if some linear combinations of $a$, $b$, and $c$ are integers,
this means that the two terms of expressions Eqs.(\ref{T_mzp1},\ref{T_z_inv}) diverge separately
when $b-a$ or $c-a-b$  respectively becomes integer, but so that their sums remain finite.
This clearly implies that Eqs.(\ref{T_mzp1},\ref{T_z_inv}) cannot be used numerically for $b-a$ or $c-a-b$  respectively
close or equal to integers due to cancellations between the two diverging terms. While it is possible to lift
analytically this indeterminacy for the integer case, the close vicinity of integers remains problematic \cite{Temme_pres,Becken_Schmelcher}.

In order to solve this problem, we will rewrite Eqs.(\ref{T_mzp1},\ref{T_z_inv}) in order to have their indeterminacy appear in a simple form, so that it can be
handled analytically, in the spirit of Ref.\cite{Forrey}. Our method is, however, much simpler than the one of Ref.\cite{Forrey},
so that it can be applied without problem to the complex case.

\section{Indeterminacy treatment in linear transformation formulas}
Two kinds of indeterminacy appear while rewriting Eqs.(\ref{T_mzp1},\ref{T_z_inv}): a few involving only elementary functions, and one the Gamma function.
The first case is handled easily using complex generalizations of the real C-functions $\mbox{expm1}(x) = \exp(x) - 1$ and $\mbox{log1p}(x) = \log(1+x)$, precise
for $|x| \sim 0$. They do not exist in Fortran 90, but can be easily coded \cite{Higham}.
The first needed functions are then:

\begin{eqnarray}
E_{\epsilon}(z) &=& \frac{\mbox{expm1}(\epsilon \; z)}{\epsilon} \mbox{ for } \epsilon \neq 0 \mbox{ , } E_{0}(z) = z \label{E_eps} , \\
P^m_{\epsilon}(z) &=& \left( \prod_{n=0 \mbox{, } n \neq n_0}^{m-1} (z + \epsilon + n) \right) \delta_{n_0 \; [0;m-1]} \nonumber \\
                  &+& (z)_m \; \frac {\displaystyle \mbox{expm1} \left[ \sum_{n=0 \mbox{, } n \neq n_0}^{m-1} \mbox{log1p} 
                      \left(\frac{\epsilon}{z+n} \right) \right]}{\epsilon}  \mbox{ for } \epsilon \neq 0 \mbox{ , } \nonumber \\
P^m_{0}(z) &=& \left( \prod_{n=0 \mbox{, } n \neq n_0}^{m-1} (z + n) \right) \delta_{n_0 \; [0;m-1]} 
               + (z)_m \sum_{n=0 \mbox{, } n \neq n_0}^{m-1} \frac{1}{z+n} ,  \\                 
\mbox{sinc}(\epsilon) &=& \frac {\sin(\pi \epsilon)}{\pi \epsilon} \mbox{ for }\epsilon \neq 0 \mbox{ , } \mbox{sinc}(0) = 1. \label{sinc}
\end{eqnarray}
where $z \in \C$, $\epsilon \in \C$, $m \in \N$, $n_0 = -| \Re(z) \rfloor$
($| x \rfloor$ is the closest integer to the real number $x$), and $\displaystyle \delta_{n_0 \; [0;m-1]}$ is equal to one if $n_0 \in [0;m-1]$ and vanishes otherwise.
The expressions provided by $P^m_{\epsilon}(z)$ with $\epsilon \neq 0$ and $P^m_{0}(z)$ are defined and stable in all cases,
as $\displaystyle |z+n| \geq \frac{1}{2}$ $\forall n \neq n_0$, immediate from previous definitions.
Note that $\displaystyle P^m_{\epsilon}(z) = \frac{(z+\epsilon)_m - (z)_m}{\epsilon}$ for $\epsilon \neq 0$.
sinc is the standard cardinal sine function, rewritten here for clarity due to its different possible definitions.

The indeterminacy involving the Gamma function is embedded in the following function:
\begin{eqnarray}
&&G_{\epsilon}(z) = \frac{1}{\epsilon} \left( \frac{1}{\Gamma}(z) - \frac{1}{\Gamma}(z+\epsilon) \right) \mbox{ , } \epsilon \neq 0 \nonumber \\
&&G_{0}(z) = \frac{\Gamma'(z)}{\Gamma(z)^2} \mbox{ , } z \in \C \backslash \Z^- \nonumber \\
&&G_{0}(z) = (-1)^{n+1} \; n! \mbox{ , } z = -n \mbox{ , } n \in \N \label{G_eps_z} .
\end{eqnarray}

The gamma function is very efficiently implemented with the Lanczos approximation \cite{NR}:
\begin{eqnarray}
\Gamma(z) \simeq \sqrt{2 \pi} \left( z + \gamma - \frac{1}{2} \right)^{z - \frac{1}{2}} e^{-\left( z + \gamma - \frac{1}{2} \right)}
                 \left[ c_0 + \sum_{i=1}^{N} \frac{c_i}{z - 1 + i} \right] \label{Lanczos_Gamma}
\end{eqnarray}
where $\displaystyle \Re(z) > 0$
and $\gamma$, $N$ and the coefficients $c_i$, $i \in [0:N]$ are chosen so $\Gamma(z)$ is precise up to numerical accuracy.
Although Eq.(\ref{Lanczos_Gamma}) is valid for $\Re(z) > 0$, it is not used close to the imaginary axis to avoid numerical instabilities.
To handle the rest of the complex plane, the Euler reflection formula $\displaystyle \Gamma(z) \Gamma (1-z) = \frac{\pi}{\sin(\pi z)}$ is applied.

When $\epsilon$ is not too close to zero, i.e.~$|\epsilon|_{\infty} > 0.1$,
a direct calculation of $G_{\epsilon}(z)$ with Eq.(\ref{Lanczos_Gamma}) is precise.
If $z \in \Z^{-}$ or $z+\epsilon \in \Z^{-}$, and $\epsilon \neq 0$, there cannot be any numerical cancellation either
as then $\displaystyle \frac{1}{\Gamma}(z) = 0$ or $\displaystyle \frac{1}{\Gamma}(z+\epsilon) = 0$,
and the case of having $z \in \Z^{-}$ and $\epsilon = 0$ is trivial.
In order to calculate $G_{\epsilon}(z)$ in the remaining case, prone to numerical error,
one introduces the following function:
\begin{eqnarray}
H_{\epsilon}(z) = G_{\epsilon}(z) \Gamma(z+\epsilon) . \label{H_eps_z}
\end{eqnarray}
The latter will be evaluated using the fact that the Lanczos approximate expression of the Gamma function is elementary.

From Eq.(\ref{Lanczos_Gamma}), one obtains for $0 < |\epsilon|_{\infty} \leq 0.1$, and $\Re(z) \geq \frac{1}{2}$ or $\Re(z + \epsilon) \geq \frac{1}{2}$:
\begin{eqnarray}
\log \frac{\Gamma (z + \epsilon)}{\Gamma(z)} &\simeq& \left( z - \frac{1}{2} \right) \mbox{log1p} \left( \frac{\epsilon}{z + \gamma - \frac{1}{2}} \right)
+ \epsilon \; \log (z + \gamma - \frac{1}{2} + \epsilon) - \epsilon \nonumber \\
&+& \mbox{log1p} \left(- \epsilon \frac{\displaystyle \sum_{i=1}^{N}
\frac{c_i}{(z - 1 + i)(z - 1 + i + \epsilon)}}
{\displaystyle c_0 + \sum_{i=1}^{N} \frac{c_i}{z - 1 + i} } \right) , \nonumber \\
H_{\epsilon}(z) &=& \frac{\mbox{expm1} \left( \log \frac{\displaystyle \Gamma (z + \epsilon)}{\displaystyle \Gamma(z)} \right)}{\epsilon} . \label{H_eps_pos}
\end{eqnarray}
Due to the well-behaved character of expm1 and log1p for very small values of $|\epsilon|_{\infty}$, 
Eq.(\ref{H_eps_pos}) provides a stable implementation of $H_{\epsilon}$.

The condition $0 < |\epsilon|_{\infty} \leq 0.1$ is still being verified,
but now for $\Re(z) < \frac{1}{2}$ and $\Re(z + \epsilon) < \frac{1}{2}$, one has to use the Euler reflection
formula along with Eq.(\ref{Lanczos_Gamma}), as well as the intermediate function $H_{-\epsilon}(z+\epsilon)$:
\begin{eqnarray}
H_{-\epsilon}(z+\epsilon) &=& \frac{1}{\epsilon} \left[ 1 - \frac{\Gamma(z)}{\Gamma(z+\epsilon)} \right] \nonumber \\
&=& \frac{1}{\epsilon} \left[ 1 - \frac{\Gamma(1-z-\epsilon)}{\Gamma(1-z)}
\left( \cos (\pi \epsilon) + \frac{\sin (\pi \epsilon)}{\tan (\pi z)} \right) \right] \nonumber \\
&=& \left( \cos (\pi \epsilon) + \frac{\sin (\pi \epsilon)}{\tan (\pi z)} \right) H_{-\epsilon}(1-z)
+ \frac{\pi^2 \epsilon}{2} \mbox{sinc}^2 \left( \frac{\epsilon}{2} \right) - \frac{\pi \mbox{ sinc}(\epsilon)}{\tan (\pi z)} , \label{H_eps_neg_intermediate} \\
H_{\epsilon}(z) &=& \frac{H_{-\epsilon}(z+\epsilon)}{1 - \epsilon \; H_{-\epsilon}(z+\epsilon)} \label{H_eps_neg}
\end{eqnarray}
where $H_{-\epsilon}(1-z)$ in Eq.(\ref{H_eps_neg_intermediate}) is calculated using Eq.(\ref{H_eps_pos})
and the Euler reflection formula has been used to obtain Eq.(\ref{H_eps_neg_intermediate}).
Clearly, Eqs.(\ref{H_eps_neg_intermediate},\ref{H_eps_neg}) can be precisely evaluated even when $|\epsilon|_{\infty}$ is very small.

The $\epsilon = 0$ case is immediate using Eq.(\ref{H_eps_pos}) and Eqs.(\ref{H_eps_neg_intermediate},\ref{H_eps_neg}) with $\epsilon \rightarrow 0$:
\begin{eqnarray}
&&H_{0}(z) \simeq \frac{z - \frac{1}{2}}{z + \gamma - \frac{1}{2}} + \log(z + \gamma - \frac{1}{2}) - 1
- \frac{\displaystyle \sum_{i=1}^{N} \frac{c_i}{(z - 1 + i)^2}}{\displaystyle c_0 + \sum_{i=1}^{N} \frac{c_i}{z - 1 + i}}
\mbox{ , } \Re(z) \geq \frac{1}{2} \label{H_eps_pos_zero} \\
&&H_{0}(z) = H_{0}(1-z) - \frac{\pi}{\tan (\pi z)} \mbox{ , } \Re(z) < \frac{1}{2} \label{H_eps_neg_zero}
\end{eqnarray}
where $H_{0}(1-z)$ is calculated using Eq.(\ref{H_eps_pos_zero}). 
It has been numerically checked that the logarithmic derivative of the Lanczos expression (see Eq.(\ref{Lanczos_Gamma})),
providing Eq.(\ref{H_eps_pos_zero}), was precise up to numerical precision to the logarithmic derivative of the Gamma function
for $z \in \C$ so that $\Re(z) \geq \frac{1}{2}$.

Note that to implement $\sin(\pi z)$ and $\tan (\pi z)$ precisely in previous formulas, one has to use the integer $n = | \Re(z) \rfloor$
and calculate instead $\sin(\pi (z-n))$ and $\tan (\pi (z-n))$ due to the finite number of digits used for $\pi$.

Finally, one arrives at the required expression for $G_{\epsilon}(z)$ for $|\epsilon|_{\infty} \leq 0.1$, with
$z \in \C \backslash \Z^{-}$ and $z+\epsilon \in \C \backslash \Z^{-}$:
\begin{eqnarray}
G_{\epsilon}(z) &=& \frac{H_{\epsilon}(z)}{\Gamma(z+\epsilon)} \mbox{ , } |z + |n||_{\infty} < |z + \epsilon + |m||_{\infty} \\
&=& \frac{H_{-\epsilon}(z+\epsilon)}{\Gamma(z)} \mbox{ , } |z + |n||_{\infty} \geq |z + \epsilon + |m||_{\infty}
\end{eqnarray}
where $n = | \Re(z) \rfloor$ and $m = | \Re(z+\epsilon) \rfloor$. 
These two equivalent equalities are used for $G_{\epsilon}(z)$ in order to have the Gamma function ratio occurring
in Eqs.(\ref{H_eps_pos},\ref{H_eps_neg_intermediate}) as small as possible in case $z$ or $z+\epsilon$ are close to a negative integer.

\section{Numerically stable linear transformation formulas}
Eqs.(\ref{T_mzp1},\ref{T_z_inv}) are now rewritten so they can be implemented without a problem. Let us recall that the polynomial case for $F(z)$
is not considered with these two formulas.

\subsection{$1-z$ transformation formula} \label{mzp1_transformation}
In order to rewrite Eq.(\ref{T_mzp1}), it is necessary for $\Re(c-a-b) \geq 0$ to be verified.
There is, however, no loss of generality, as one can apply Eq.(\ref{T_cab}) in case $\Re(c-a-b) < 0$ to reach the required condition.
One defines the integer $m = | \Re(c-a-b) \rfloor$ and the complex number $\epsilon = c - a - b - m$.
One then obtains for Eq.(\ref{T_mzp1}), after some tedious manipulations,

\begin{eqnarray}
F(z) &=& \frac{(-1)^m}{\mbox{sinc}(\epsilon)} \left( A(z) + B(z) \right) , \\
A(z) &=& \frac{\Gamma(c)}{\epsilon \; \Gamma(c-a) \Gamma(c-b)} \sum_{n=0}^{m-1} \frac{(a)_n (b)_n}{n! \; \Gamma(1+n-m-\epsilon)} (1-z)^n , \label{Az_mzp1} \\
B(z) &=& \frac{\Gamma(c)}{\epsilon \; (a+\epsilon)_m (b+\epsilon)_m}
\sum_{n=0}^{+\infty} \left[ \frac{(a)_{n+m} (b)_{n+m}}{\Gamma(a+\epsilon) \Gamma(b+\epsilon) \Gamma(n+m+1) \Gamma(n+1-\epsilon)} \right. \nonumber \\
&-& \left. \frac{(1-z)^{\epsilon} \; (a+\epsilon)_{n+m} (b+\epsilon)_{n+m}}{\Gamma(a) \Gamma(b) \Gamma(n+m+1+\epsilon) \Gamma(n+1)} \right] (1-z)^{n+m}
\label{Bz_mzp1}
\end{eqnarray}
where $z \in \C$ so that $|1-z| < 1$ and $\epsilon \in \C^*$. $\epsilon = 0$ will be considered afterward as a limiting case.

Let us first consider the implementation of the finite sum $A(z)$ of Eq.(\ref{Az_mzp1}).
If one writes $\displaystyle A(z) = \sum_{n=0}^{m-1} \alpha_n (1-z)^n$, one obtains:
\begin{eqnarray}
\alpha_0 &=& \frac{\Gamma(c)}{\epsilon \; \Gamma(1-m-\epsilon) \Gamma(a + m + \epsilon) \Gamma (b + m + \epsilon)} \mbox{ , } \epsilon \neq 0 \nonumber \\
&=& (-1)^m (m-1)! \frac{\Gamma(c)}{\Gamma(a + m) \Gamma (b + m)} \mbox{ , } \epsilon = 0 , \label{a0_mzp1} \\
\alpha_{n+1} &=& \frac{(a+n)(b+n)}{(n+1)(1-m-\epsilon+n)} \alpha_n \mbox{ , } 0 \leq n \leq m-2 .
\end{eqnarray}

In order to implement $B(z)$, one writes $\displaystyle B(z) = \sum_{n=0}^{+\infty} \beta_n (1-z)^n$, and using Eq.(\ref{Bz_mzp1}) one obtains
\begin{eqnarray}
\beta_0 &=& \left[ \frac{(a)_m (b)_m}{\Gamma(1-\epsilon) \Gamma(a + m + \epsilon) \Gamma (b + m + \epsilon) \Gamma(m+1)}
- \frac{(1-z)^\epsilon}{\Gamma(a) \Gamma(b) \Gamma(m+1+\epsilon)} \right] \nonumber \\
&\times& \frac{\Gamma(c) (1-z)^m}{\epsilon} \mbox{ , } |\epsilon|_{\infty} > 0.1  \nonumber \\
&=& \left[ \frac{1}{\Gamma(a + m + \epsilon) \Gamma (b + m + \epsilon)} \left( \frac{G_{-\epsilon}(1)}{\Gamma(m+1)} + G_{\epsilon}(m+1) \right) \right.
\nonumber\\
&-& \frac{1}{\Gamma(m + 1 + \epsilon)} \left( \frac{G_{\epsilon}(a+m)}{\Gamma(b+m+\epsilon)} + \frac{G_{\epsilon}(b+m)}{\Gamma(a+m)} \right)
\nonumber\\
&-& \left. \frac{E_{\epsilon}(\log(1-z))}{\Gamma(a + m) \Gamma (b + m) \Gamma(m+1+\epsilon)} \right]
\times \Gamma(c) (a)_m (b)_m (1-z)^m  \mbox{ , } |\epsilon|_{\infty} < 0.1 , \label{b0_mzp1} \\
\gamma_0 &=& \frac{\Gamma(c) (a)_m (b)_m (1-z)^m}{\Gamma(a + m + \epsilon) \Gamma (b + m + \epsilon) \Gamma(m+1) \Gamma(1 - \epsilon)} , \label{g0_mzp1} \\
\beta_{n+1} &=& \frac{(a+m+n+\epsilon)(b+m+n+\epsilon)}{(m+n+1+\epsilon)(n+1)} \beta_n
+ \left[ \frac{(a+m+n)(b+m+n)}{(m+n+1)} \nonumber \right. \\
&-& \left. (a+m+n) - (b+m+n) - \epsilon + \frac{(a+m+n+\epsilon)(b+m+n+\epsilon)}{n+1} \right] \nonumber \\
&\times&    \frac{\gamma_n}{(n+m+1+\epsilon)(n+1-\epsilon)} \mbox{ , } n \geq 0 , \\
\gamma_{n+1} &=& \frac{(a+m+n)(b+m+n)}{(n+m+1)(n+1-\epsilon)} \gamma_n \mbox{ , } n \geq 0
\end{eqnarray}
where $\gamma_n$ is an auxiliary term needed for the calculation of $\beta_n$.

One can see that the expressions defining $A(z)$ and $B(z)$ through their respective series terms $\alpha_n$ and $\beta_n$ are advantageous for numerical
computation. Firstly, Gamma and elementary function evaluations only occur in the first term of the series,
while all other are rational functions of their predecessors. This implies a fast evaluation of the series.
Secondly, the complication induced by the treatment of small $|\epsilon|_{\infty}$ is also present only at the level of the first
term of $A(z)$ and $B(z)$, so that it is not prohibitive to be in the small $|\epsilon|_{\infty}$ regime compared to the normal regime.

\subsection{$\displaystyle \frac{1}{z}$ transformation formula} \label{z_inv_transformation}
The method used for the transformation in Eq.(\ref{T_z_inv}) is very similar to the one of Sec.(\ref{mzp1_transformation}).
In this case, the relation $\Re(b-a) \geq 0$ must be verified. Here again, there is no loss of generality, 
as Eq.(\ref{T_ba}) can be applied in the case $\Re(b-a) < 0$.
We will use the same notations as in Sec.(\ref{mzp1_transformation}), except for $m = | \Re(b-a) \rfloor$ and $\epsilon = b - a - m$.
One then has for Eq.(\ref{T_z_inv}):

\begin{eqnarray}
F(z) &=& \frac{(-1)^m (-z)^{-a}}{\mbox{sinc}(\epsilon)} \left( A(z) + B(z) \right) , \label{hyp_one_over_z} \\
A(z) &=& \frac{\Gamma(c)}{\epsilon \; \Gamma(b) \Gamma(c-a)} \sum_{n=0}^{m-1} \frac{(a)_n (1-c+a)_n}{\Gamma(n+1) \Gamma(1+n-m-\epsilon)} z^{-n}
, \label{Az_z_inv} \\
B(z) &=& \frac{\Gamma(c)}{\epsilon} \sum_{n=0}^{+\infty} \left[ \frac{(a)_{n+m} (1-c+a)_{n+m}} {\Gamma(a+\epsilon) \Gamma(c-a) \Gamma(n+m+1) \Gamma(n+1-\epsilon)
(a+\epsilon)_m (b+\epsilon)_m} \right. \nonumber \\
&-& \left. \frac{(-z)^{-\epsilon} \; (a+\epsilon)_{n+m} (1-c+a+\epsilon)_{n+m}}{(a+\epsilon)_m \; \Gamma(a)
\Gamma(c-a-\epsilon) \Gamma(n+m+1+\epsilon) \Gamma(n+1)} \right] z^{-(n+m)} \label{Bz_z_inv}
\end{eqnarray}
where $z \in \C$ so that $|z| > 1$ and $\epsilon \in \C^*$.

Their series terms read:
\begin{eqnarray}
\alpha_0 &=& \frac{\Gamma(c)}{\epsilon \; \Gamma(1-m-\epsilon) \Gamma(a+m+\epsilon) \Gamma (c-a)} \mbox{ , } \epsilon \neq 0 \nonumber \\
&=& (-1)^m (m-1)! \frac{\Gamma(c)}{\Gamma(a+m) \Gamma (c-a)} \mbox{ , } \epsilon = 0 , \label{a0_z_inv} \\
\alpha_{n+1} &=& \frac{(a+n)(1-c+a+n)}{(n+1)(1-m-\epsilon+n)} \alpha_n \mbox{ , } 0 \leq n \leq m-2 , \\
\beta_0 &=& \frac{\Gamma(c) z^{-m}}{\epsilon} \left[ \frac{(a)_m (1-c+a)_m}{\Gamma(c-a) \Gamma(a+m+\epsilon) \Gamma(1-\epsilon) \Gamma(m+1)} \right. \nonumber \\
&-& \left. \frac{(1-c+a+\epsilon)_m (-z)^{-\epsilon}}{\Gamma(a) \Gamma(c-a-\epsilon) \Gamma(m+1+\epsilon)} \right]
\mbox{ , } |\epsilon|_{\infty} > 0.1 \nonumber \\
&=& (a)_m \Gamma(c) z^{-m} \left[ \frac{(1-c+a+\epsilon)_m G_{-\epsilon}(1) - P^m_{\epsilon}(1-c+a) \frac{\displaystyle 1}{\displaystyle \Gamma}(1-\epsilon)}
{\Gamma(c-a) \Gamma(a+m+\epsilon) \Gamma(m+1)} \right. \nonumber \\
&+& (1-c+a+\epsilon)_m \left[ \frac{G_{\epsilon}(m+1)}
{\displaystyle \Gamma(c-a) \Gamma(a+m+\epsilon)} - \frac{G_{\epsilon}(a+m)}{\Gamma(c-a) \Gamma(m+1+\epsilon)} \right. \nonumber \\
&-& \left. \left. \frac{G_{-\epsilon}(c-a) 
- E_{-\epsilon}(\log(-z)) \frac{\displaystyle 1}{\displaystyle \Gamma}(c-a-\epsilon)}{\Gamma(m+1+\epsilon) \Gamma(a+m)} \right] \right]
\mbox{ , } |\epsilon|_{\infty} < 0.1 , \label{b0_z_inv} \\
\gamma_0 &=& \frac{(a)_m (1-c+a)_m \Gamma(c) z^{-m}}{\Gamma(a+m+\epsilon) \Gamma(c-a) \Gamma(1-\epsilon) \Gamma(m+1)} , \label{g0_inv} \\
\beta_{n+1} &=& \frac{(a+m+n+\epsilon)(1-c+a+m+n+\epsilon)}{(m+n+1+\epsilon)(n+1)}\beta_n \nonumber \\
&+& \left[ \frac{(a+m+n)(1-c+a+m+n)}{(m+n+1)} - (a+m+n) \right. \nonumber \\
&-& \left. (1-c+a+m+n) - \epsilon + \frac{(a+m+n+\epsilon)(1-c+a+m+n+\epsilon)}{n+1} \right] \nonumber \\
&\times& \frac{\gamma_n}{(n+m+1+\epsilon)(n+1-\epsilon)} \mbox{ , } n \geq 0 , \\
\gamma_{n+1} &=& \frac{(a+m+n)(1-c+a+m+n)}{(n+m+1)(n+1-\epsilon)} \gamma_n \mbox{ , } n \geq 0 .
\end{eqnarray}

Obviously, this scheme has the same numerical advantages as the $1-z$ transformation formula of Sec.(\ref{mzp1_transformation}).

\section{Power series convergence test} \label{PS_cv_test}
In the formulas of Sec.(\ref{linear_transf_section}), the power series defined in Eq.(\ref{hyp_PS_zero}) is used with various sets of parameters.
Due to the presence of Pochhammer symbols, its general term modulus may increase as $n$ goes to $+\infty$, before finally going to zero.
The problem to tackle is the possible appearance of false convergence. If one truncates the power series when its
general term modulus is smaller than a given precision, one may obtain a wrong or inaccurate value if it eventually appears to increase afterward.
In Ref.\cite{Forrey}, the rest of the power series was majored by an analytical series in order to evaluate the needed number of terms for convergence.
In order to avoid a false convergence, it is not necessary, but sufficient, 
to know whether the modulus of the general term of Eq.(\ref{hyp_PS_zero}) is either increasing or decreasing to zero for $n$ large enough.

For that, one defines a polynomial $P(x)$ built upon the ratio of the modulus of two consecutive general terms (see  Eq.(\ref{hyp_PS_zero_term})):
\begin{eqnarray}
P(x) = \left| z \; (a+x)(b+x) \right|^2 -  \left| (c+x)(x+1) \right|^2
\end{eqnarray}
which is clearly a polynomial of degree four in $x$ when $a$, $b$, $c$, and $z$ are fixed.
$P(n)$ has the property to be positive when the general term of Eq.(\ref{hyp_PS_zero}) increases, while it is negative when it decreases.
The knowledge of the sign of $P(x)$ would demand the calculation of its four roots, which would be too lengthy even with analytical formulas.
It is sufficient, however, to consider its first and second derivatives, the latter being a polynomial of degree two. Indeed, with $\Delta$
the discriminant of $P''(x)$, $\Delta \leq 0$ implies that $P(x)$ has only one maximum while $\Delta > 0$ means that it has two local maxima and one local minimum.
If $\Delta \leq 0$ and $P'(x_0) < 0$ for a real $x_0$, one will have $P'(x) < 0$ $\forall x \geq x_0$. Thus, false convergence becomes impossible to occur
once an integer $n_0$ verifying $P'(n_0) < 0$ is reached. If $\Delta > 0$, one considers the largest real root $x_c$ of $P''(x)$, and $n_c$ the smallest integer
larger or equal to $x_c$. The curvature of $P(x)$ becoming negative for $x > x_c$, $P(x)$ has at most one maximum in this zone.
Hence, a similar property is obtained as for $\Delta \leq 0$, that if $P'(x_0) < 0$ for $x_0 > x_c$, $P'(x) < 0$ $\forall x \geq x_0$,
so that false convergence cannot appear with integers $n > n_c$.
Consequently, during the power series evaluation, it is sufficient to calculate $P'(n)$ as long as $P'(n) \geq 0$ for $n > 0$ ($\Delta \leq 0$) or
for $n \geq n_c$ ($\Delta > 0$), and, when $P'(n) < 0$,
to check if its general term at integer $n$ is smaller than numerical precision. There, the power series can be truncated safely.
As this condition arrives in general very quickly, this method does not add expensive additional test calculations which would slow the power series
evaluation. If Eqs.(\ref{T_mzp1},\ref{T_z_inv}) are used, this test is considered for the two power series present in these transformations.

\section{Numerical algorithm} \label{hyp_2F1_algo}

We consider first the non-polynomial case, implying that $\displaystyle c \in \C \backslash \Z^{-}$.

\subsection{General case}
The cut in $[1:+\infty[$ is handled replacing $z \geq 1$ by $z + i0^-$ (limit from below the real axis),
so that $z$ becomes $z - i 10^{-307}$ in the code.
One begins with applying the linear transformation of Eq.(\ref{T_ba}) if $\Re(b-a) < 0$ and the one of Eq.(\ref{T_cab}) if $\Re(c-a-b) < 0$
in order to be able to to use Eqs.(\ref{T_mzp1},\ref{T_z_inv}). The latter transformation also has the advantage to accelerate Eq.(\ref{hyp_PS_zero}) convergence,
as its term behaves like $\displaystyle \frac{z^n}{n^{c-a-b}}$ for $n \rightarrow +\infty$, 
easy to demonstrate using Stirling's approximation for Pochhammer symbols and factorial.

If $|1-z| < 10^{-5}$, the transformation of Eq.(\ref{T_mzp1}) with the method of Sec.(\ref{mzp1_transformation}) is applied,
as it handles the vicinity of the singular point $z=1$ properly.

Then, the $|z|$ and  $\displaystyle \left| \frac{z}{z-1} \right|$ moduli are compared respectively to a given radius $R$,
which runs from 0.5 to 0.9. $R = 0.9$ may appear seemingly large, but it was acknowledged that power series summations are still fast and precise
in this regime if $|a|$, $|b|$ and $|c|$ are small or moderate.
Moreover, although the number of terms necessary to converge may be smaller using Eqs.(\ref{T_mzp1},\ref{T_z_inv}),
this advantage is hindered by the larger number of multiplications and divisions per series term as well as the presence of several Gamma function evaluations.

If $|z|$ or $\displaystyle \left| \frac{z}{z-1} \right|$ happens to be smaller than $R$,
Eq.(\ref{hyp_PS_zero}) or Eqs.(\ref{hyp_PS_zero},\ref{T_ma}) are processed, respectively.
Other transformations are considered thereafter if $|z| > 0.9$ and $\displaystyle \left| \frac{z}{z-1} \right| > 0.9$.

{The $\displaystyle {\left| \frac{1}{z} \right|}$, $\displaystyle {\left| \frac{z-1}{z} \right|}$, ${|1-z|}$, 
and $\displaystyle {\left| \frac{1}{1-z} \right|}$
moduli are compared respectively to $R$, which runs again from 0.5 to 0.9.} If one of them is found to be smaller than $R$, the corresponding transformation,
i.e.~the one using Eq.(\ref{T_z_inv}), Eqs.(\ref{T_ma},\ref{T_z_inv}), Eq.(\ref{T_mzp1}), or Eqs.(\ref{T_ma},\ref{T_mzp1}), respectively, is applied.
Note that the conditions $\Re(b-a) \geq 0$ and $\Re(c-a-b) \geq 0$ are always verified even when the transformation of Eq.(\ref{T_ma}) is utilized.
Eq.(\ref{T_mzp1}), however, is considered only if $|a|_{\infty} < 5$, $|b|_{\infty} < 5$, and $|c|_{\infty} < 5$.
Indeed, the transformation of Eq.(\ref{T_mzp1}) has been found to become quickly unstable when $|a|,|b|$, and $|c|$ increase.
Transformations involving Eq.(\ref{T_ma}) are also processed only if $|c-b|_{\infty} < 5$ in order to avoid instabilities.
The number of terms needed to reach convergence with the power series is determined with the method explained in Sec.(\ref{PS_cv_test}).

If none of the aforementioned transformations succeed, it has been chosen to determine $F(z)$ through a Taylor series expansion.
Indeed, the missing zones of our linear transformation method, located in the vicinity of $\displaystyle z = e^{\pm i \frac{\pi}{3}}$, 
verify $0.9 < |z| < 1.1$, so that for small or moderate $|a|$, $|b|$, and $|c|$,
they are close enough to accessible regions to be tackled with such an expansion without inducing instabilities.
For this, one defines the center of the Taylor expansion disk $\displaystyle z_0 = r_0 \frac{z}{|z|}$,
where $r_0=0.9$ if $|z| < 1$ and $r_0=1.1$ if $|z| > 1$. One then has:
\begin{eqnarray}
&&F(z) = \sum_{n=0}^{+\infty} q_n (z-z_0)^n \label{Taylor_rest}
\end{eqnarray}
with:
\begin{eqnarray}
&&q_0 = F(z_0) \mbox{ , } q_1 = F'(z_0) = \frac{ab}{c} {_2F_1}(a+1,b+1,c+1;z_0) , \nonumber \\
&&q_{n+2} = \frac{1}{z_0(1-z_0)(n+2)} \left[ \left( n(2 z_0 - 1) - c + (a+b+1) z_0 \right) q_{n+1} + \frac{(a+n)(b+n)}{n+1} q_n \right], \nonumber \\
&&n \geq 0 \label{Taylor_rest_iter_coeff}
\end{eqnarray}
where $q_0$ and $q_1$ are calculated with Eq.(\ref{hyp_PS_zero}) if $r_0 = 0.9$, and with Eq.(\ref{hyp_one_over_z}) if $r_0 = 1.1$.
No expression in denominators may induce false convergence in the three-term recurrence relation of Eq.(\ref{Taylor_rest_iter_coeff}),
contrary to Eq.(\ref{hyp_PS_zero_term}), where $c+n$ can become very small and then make the general term of the series increase abruptly
in modulus. On the contrary, here $q_n$ behaves smoothly, so that testing convergence with $|q_n (z-z_0)^n|_{\infty} + |q_{n+1} (z-z_0)^{n+1}|_{\infty}$,
to avoid eventual single instances of very small $q_{n}$, is sufficient. As $|z - z_0| \leq 0.1$, the Taylor series of Eq.(\ref{Taylor_rest})
converges very quickly, with 10-20 terms typically. Hence, virtually all computational time will be spent calculating $F(z_0)$ and $F'(z_0)$,
for which $z_0$ lies close to the circle of convergence. The zones in the vicinity of $\displaystyle z = e^{\pm i \frac{\pi}{3}}$ thus demand
twice as much time as in the general case. This technique has been compared to the scheme devised in Ref.\cite{Buhring} for $b-a$ not close to integer.
Indeed, the latter demands that $b-a$ is not a strictly positive integer, and hence not a complex number very close to an integer as well.
$z$ is taken to be equal to $\displaystyle r e^{i \frac{\pi}{3}}$, with $r = 0.99$ or $r = 1.01$, so that $F(z_0)$ and $F'(z_0)$ are implemented
with respectively Eq.(\ref{hyp_PS_zero}) and Eq.(\ref{hyp_one_over_z}). Results are shown in Tabs.(\ref{table_Buhring_test_1},\ref{table_Buhring_test_2}).
If $r = 0.99$, both the method of Ref.\cite{Buhring} and ours have comparable times of calculations, respectively $\sim$ 8 seconds and $\sim$ 10 seconds to calculate Tab.(\ref{table_Buhring_test_1}), because Eq.(\ref{hyp_PS_zero}) is very quickly summed and no Gamma functions are evaluated therein.
{However, the method of Ref.\cite{Buhring} is faster for ${r = 1.01}$ by a factor of 4-5,  as it also takes ${\sim}$ 8 seconds to calculate Tab.(\ref{table_Buhring_test_1}) with it,
whereas it takes ${\sim}$ 35 seconds with Eqs.(\ref{Taylor_rest},\ref{Taylor_rest_iter_coeff}).} Nevertheless, it is clear from Tabs.(\ref{table_Buhring_test_1},\ref{table_Buhring_test_2}) 
that our method is more robust when $|a|$, $|b|$ and $|c|$ are not small. Hence, due to its ability to treat $b-a$ equal or close to integers properly
and due to its robustness, our method is to be preferred to that of Ref.\cite{Buhring}, even if it is slower for $|z| > 1$.

\subsection{Polynomial case}
If $F(z)$ is a polynomial, either $a \in \Z^{-}$ or $b \in \Z^{-}$.
If $a \in \Z^{-}$, one uses Eq.(\ref{hyp_PS_zero}) if $\displaystyle |z| < \left| \frac{z}{z-1} \right|$ and
Eqs.(\ref{hyp_PS_zero},\ref{T_ma}) if not. As both related power series terminate, it is possible to use them with $|z| > 1$.
Indeed, the power series argument modulus cannot exceed two, so that the evaluation of the polynomials remains stable. It was checked that for
$a$, $b$, and $c$ parameters of a given order of magnitude, 
this method provides the same or better precision as the more involved non-polynomial case (see Sec.(\ref{recommendations})).
If $b \in \Z^{-}$, the procedure is the same, using Eq.(\ref{T_mb}) instead of Eq.(\ref{T_ma}).

\section{Application: wave functions of the PTG potential}
Potentials sustaining analytical wave function solutions of the Schr{\"o}dinger equation are not often.
Among them, the class of Natanzon potentials \cite{Natanzon_pot1} is widely studied as they depend on six parameters,
so that they can accommodate various shapes. The PTG potential belongs to a subclass
of the Natanzon potentials and depends on four parameters. Its interest is twofold: firstly, parameters can be selected
so that it develops a flat bottom, a feature appreciated in nuclear theory as it models the interior of nuclei
(see Refs.\cite{Karim,Mario1,Mario2} for examples of applications). Secondly, it possesses a quasi-centrifugal barrier,
behaving properly for $r \rightarrow 0$, although vanishing exponentially quickly for $r \rightarrow +\infty$,
as well as the possibility to bear an effective mass.

\subsection{PTG Schr{\"o}dinger equation} \label{PTG_eq_section}
The Schr{\"o}dinger equation allowing for the PTG potential reads \cite{PTG_pot}:
\begin{eqnarray}
\left[ \frac{\hbar^2}{2 m_0} \left( - \frac{d}{dr} \frac{1}{\mu(r)} \frac{d}{dr} + \frac{\ell(\ell+1)}{r^2 \mu(r)} \right) + V_{PTG}(r) \right] u(r) = e \; u(r)
\label{PTG_equation}
\end{eqnarray}
with $m_0$ the particle free mass, $\mu(r)$ its dimensionless effective mass (the full effective mass is $m_0 \; \mu(r)$), $\ell$ its orbital angular momentum,
$V_{PTG}(r)$ the PTG potential, detailed in this section, $u(r)$ its wave function, and $e$ its energy.
$r$ has the dimension of a length and $m_0 c^2$ ($c$ is the speed of light), as well as $e$, of an energy.

$V_{PTG}$ depends on four parameters, conventionally denoted as $\Lambda$, $s$, $\nu$, and $a$. $\Lambda$ controls the shape of the potential so that
it is very diffuse for small $\Lambda$, becoming the usual P{\"o}schl-Teller potential for $\Lambda = 1$, while for larger values of $\Lambda$, a flat bottom
develops. $s$ is a scaling parameter as $V_{PTG}$ is proportional to its square, and $\nu$ determines the depth of the potential. Finally, $a$ wields
the strength of the effective mass and varies between 0 and 1, $(1-a)m_0$ being the effective mass at $r = 0$, while the free effective mass $m_0$ is 
recovered for $r \rightarrow +\infty$.

$\mu(r)$ and $V_{PTG}(r)$ are expressed through the variable $y$ depending on $r$, defined by way of an implicit equation \cite{PTG_pot}:
\begin{eqnarray}
\Lambda^2 s~ r &=& \mbox{arctanh}(y) + \sqrt{\Lambda^2 - 1} \mbox{ } \arctan(\sqrt{\Lambda^2 - 1} \; y) \mbox{ , } \Lambda > 1 \label{y_equation_1} \\
&=& \mbox{arctanh}(y) - \sqrt{1 - \Lambda^2} \mbox{ arctanh}(\sqrt{1 - \Lambda^2} \; y) \mbox{ , } \Lambda \leq 1 . \label{y_equation_2}
\end{eqnarray}
A numerical technique to quickly and precisely solve this equation will be delineated afterward.

$V_{PTG}(r)$ is separated in three parts: $V_{\mu}(r)$, directly proportional to $a$, $V_{\ell}(r)$ its $\ell$-dependent part,
and $V_{c}(r)$, its principal central part. One can now outline the expressions for $\mu(r)$ and $V_{PTG}(r)$ \cite{PTG_pot}:
\begin{eqnarray}
\mu(r) &=& 1 - a(1-y^2) , \label{mu_r} \\
V_{\mu}(r) &=& \left[ 1-a + \left[ a(4-3 \Lambda^2) - 3(2 - \Lambda^2) \right] y^2 - (\Lambda^2 - 1)(5(1-a) + 2 a y^2) \; y^4 \right] \nonumber \\
&\times& \frac{a}{\mu(r)^2} (1-y^2) \left[ 1 + (\Lambda^2 - 1) y^2 \right] , \\
V_{\ell}(r) &=& \ell(\ell+1) \left[ \frac{(1 - y^2)( 1 + (\Lambda^2 - 1) y^2)}{y^2} - \frac{1}{s^2 r^2} \right]
\mbox{ , } r > 0 \label{Vl_r} , \\
V_{c}(r) &=& (1-y^2) \left[ -\Lambda^2 \nu (\nu+1) - \frac{\Lambda^2 - 1}{4}
\left( 2 - (7 - \Lambda^2) y^2 - 5(\Lambda^2 - 1) y^4 \right)  \right] , \\
V_{PTG}(r) &=& \frac{\hbar^2 s^2}{2 m_0 \mu(r)} \left( V_{\mu}(r) + V_{\ell}(r) + V_{c}(r) \right) \label{V_PTG}
\end{eqnarray}

The expression of $V_{\ell}(r)$ in Eq.(\ref{Vl_r}) presents numerical cancellations for $r \rightarrow 0$,
as $y \sim sr$ therein so that both terms of Eq.(\ref{Vl_r}) diverge but their sum remains finite.
Divergence is suppressed if one employs Eqs.(\ref{y_equation_1},\ref{y_equation_2})
to rewrite Eq.(\ref{Vl_r}) so that Eq.(\ref{Vl_r}) can be evaluated via a power series:
\begin{eqnarray}
V_{\ell}(r) &=& \ell(\ell+1) \left[ \frac{y}{\Lambda^2sr} \left( 1 + \frac{y}{sr} \right) S_{\Lambda}(y)
+ \Lambda^2 - 1 - (1 + (\Lambda^2 - 1) y^2)  \right] \label{stable_Vl_r} \mbox{ , } r > 0 , \\
S_{\Lambda}(y) &=& \sum_{n=0}^{+\infty} \frac{1 - (1-\Lambda^2)^{n+2}}{2n+3} y^{2n} , \label{PS_for_Vl} \\
V_{\ell}(0) &=& \ell(\ell+1) \left( \frac{\Lambda^2 - 2}{3} \right) \label{Vl_0} .  
\end{eqnarray}
Eq.(\ref{stable_Vl_r}) will be used when $r>0$ and parameters of arctanh and $\arctan$ in Eqs.(\ref{y_equation_1},\ref{y_equation_2}) are smaller than 0.01, so
that Eq.(\ref{PS_for_Vl}) will converge quickly. $V_{\ell}(0)$ is analytical and provided by Eq.(\ref{Vl_0}).
With this method, virtually full accuracy is obtained for $V_{\ell}(r) \; \forall r \geq 0$.

\subsection{y(r) function calculation} \label{y_calculation}

The parameter $y(r)$ defined in Eqs.(\ref{y_equation_1},\ref{y_equation_2})
has to be handled numerically, as it will be expressed analytically only for the P{\"o}schl-Teller potential
with $\Lambda = 1$, where it is equal to $\tanh(\Lambda^2 s r)$. Two problems appear while solving this equation. The first one is related to the rapid
variations of $\mbox{arctanh}(y)$ for $y \sim 1$, which occurs when $r$ increases. The second problem comes from the saturation of $y$ to the value $y = 1$
for $r \rightarrow +\infty$.  When $y \sim 1$ then $1 - y^2$ quickly becomes inaccurate by a direct computation.
A precise knowledge of $1 - y^2$ is indeed demanded to calculate properly $\mu(r)$ and $V_{PTG}(r)$
(see Eqs.(\ref{mu_r},\ref{V_PTG})).  

Although $y(r)$ is not analytical, it is possible to locate it analytically inside the $[0:1[$ interval. Indeed, $y$ belongs to the interval $[y_d:y_e]$,
where $y_d$ and $y_e$ read:
\begin{eqnarray}
y_d &=& \max (\tanh(\Lambda^2 s r - \sqrt{\Lambda^2 - 1} \mbox{ } \arctan(\sqrt{\Lambda^2 - 1})),0) \mbox{ , } \Lambda > 1 \\
&=& \tanh(\Lambda^2 s r) \mbox{ , } \Lambda^2 \leq 1  . \\
y_e &=& \tanh(\Lambda^2 s r) \mbox{ , } \Lambda^2 > 1  \\
&=& \tanh(\Lambda^2 s r + \sqrt{1 - \Lambda^2} \mbox{ arctanh}(\sqrt{1 - \Lambda^2})) \mbox{ , } \Lambda \leq 1 .
\end{eqnarray}

Moreover, $y(r)$ verifies simple expansions for $y \sim 0$ and $y \sim 1$:
\begin{eqnarray}
y(r) &=& sr + O(y^3) \mbox{ , } r \rightarrow 0 \label{y_r_small} \\
&=& y_d + O \left( (1-y)^2 \right) \mbox{ , } r \rightarrow +\infty \mbox{ , } \Lambda > 1 \label{y_r_large_1} \\
&=& y_e + O \left( (1-y)^2 \right) \mbox{ , } r \rightarrow +\infty \mbox{ , } \Lambda \leq 1 . \label{y_r_large_2}
\end{eqnarray}

As the behavior of $y$ at the extremities of $[0:1[$ is known, it is possible to have a quickly converging Newton method procedure to find $y(r)$,
as one can there find a very good starting point, and also because the derivative, according to $y$, of Eqs.(\ref{y_equation_1},\ref{y_equation_2}) is analytical.
One uses the following starting point $y_{start}$ for the Newton method:
\begin{eqnarray}
y_{start} &=& y_d \mbox{ if } y_d > \frac{1}{2} \mbox{ and } \Lambda > 1  \\
&=& y_e \mbox{ if } y_e > \frac{1}{2} \mbox{ and } \Lambda \leq 1  \\
&=& sr \mbox{ } (0.99 \mbox{ if } sr > 0.99) \mbox{ for other cases .}
\end{eqnarray}
It is complemented by the bisection of $[y_d:y_e]$ for the rare cases where $y$ falls out of the interval $[0:1[$ during the Newton procedure.

This procedure, however, cannot be used if $y \sim 1$ (in the code, the condition $y_d > 0.99$ is employed), as numerical inaccuracies would occur.
Hence, another method has to be employed. For this, one uses an iterative fixed-point scheme on $x = \mbox{arctanh}(y)$:
\begin{eqnarray}
x_n &=& \Lambda^2 s r - \sqrt{\Lambda^2 - 1} \mbox{ } \arctan(\sqrt{\Lambda^2 - 1} y_n)  \mbox{ , } \Lambda > 1  \\
&=& \Lambda^2 s r + \sqrt{1 - \Lambda^2} \mbox{ arctanh}(\sqrt{1 - \Lambda^2} y_n) \mbox{ , } \Lambda \leq 1 , \\
y_{n+1} &=& \tanh(x_n) \mbox{ , } n \geq 0 ,
\end{eqnarray}
and $y_0 = 1$ as a starting point.
As $y \sim 1$, $x_n$ converges quickly to $x$ and $y_n$ to $y$ for $n \rightarrow +\infty$.
This representation allows a precise determination of $1-y^2$, as $\displaystyle 1 - y^2 = \frac{4 e^{-2x}}{(1 + e^{-2x})^2}$,
expression not subject to numerical inaccuracy.

\subsection{PTG wave functions}

\subsubsection{Bound, antibound, and resonant states momenta and energies}
{The PTG potential sustains bound, antibound, and resonant states, i.e.~S-matrix poles, whose energies are known analytically \cite{PTG_pot}.
The nature of the state depends on the value of the integer ${N = 2n + \ell + 1}$, where ${n}$ is its principal quantum number,
and of the following boundary values:}
\begin{eqnarray}
B_{\Lambda} &=& \Lambda \sqrt{\frac{1-a}{\Lambda^2(1-a) - 2}} \left( \nu + \frac{1}{2} \right) - \frac{1}{2}  \mbox{ , }  \Lambda > \sqrt{\frac{2}{1-a}}  \\
&=& +\infty  \mbox{ , } \Lambda \leq \sqrt{\frac{2}{1-a}},  \\
B_{\nu} &=& [\nu] \mbox{ , } \nu \in \R \backslash \N , \\
B_{\nu} &=& \nu - 1 \mbox{ , } \nu \in \N .
\end{eqnarray}

{Bound states occur when ${N \leq B_{\Lambda}}$ and ${N \leq B_{\nu}}$, antibound states for ${N \leq B_{\Lambda}}$ and ${N > B_{\nu}}$ (${N > \nu}$ if ${\nu \in \N}$),
and resonant states for ${N > B_{\Lambda}}$. Their linear momenta and energy read:}
\begin{eqnarray}
k &=& i s \frac{-\left( N + \frac{1}{2}\right) \pm \sqrt{\Delta}}{1-a} , \\
e &=& \frac{\hbar^2 k^2}{2 m_0}
\end{eqnarray}
{where the ${\pm}$ sign is ``+'' for bound and antibound states whereas it is ``-'' for resonant or complex virtual states,
and ${\Delta}$ reads:}
\begin{eqnarray}
\Delta &=& \Lambda^2 \left( \nu + \frac{1}{2} \right)^2 (1-a) - \left[ (1-a)\Lambda^2 - 1 \right] \left[ N + \frac{1}{2}\right]^2 .
\end{eqnarray}

$N = \nu$ is suppressed for $\nu \in \N$ in the bound/antibound case as it provides the value $k = 0$, for which it is easy to show that the PTG
wave function vanishes identically.

\subsubsection{Wave functions analytical expressions} \label{PTG_wf_expression}
The general expression of PTG wave functions is very similar for S-matrix poles and scattering states \cite{PTG_pot}:
\begin{eqnarray}
\varphi(r) &=& \mathcal{N} \; \chi(r) X^+(r) F(r) \label{PTG_wf1} \\
&=& \mathcal{N} \; \chi(r) \left( A^+ X^+(r) F^+(r) + A^- X^-(r) F^-(r) \right) \label{PTG_wf2}
\end{eqnarray}
{where ${\mathcal{N}}$ is a normalization constant depending on the nature of the state and used functions read:}
\begin{eqnarray}
&&{X^+(r) = (x^+)^{\frac{\bar{\beta}}{2}}}
\mbox{ , } {X^-(r) = (x^+)^{-\frac{\bar{\beta}}{2}}}
\mbox{ , } {\chi(r) = \sqrt{\frac{x^- + \Lambda^2(1-a)x^+}{\sqrt{x^- + \Lambda^2 x^+}}} (x^-)^{\frac{\ell+1}{2}}} , \\
&&F(r) = {_2F_1} \left( \nu^-,\nu^+,\ell + \frac{3}{2} ; x^- \right) , \label{F_PTG} \\
&& F^+(r) = {_2F_1} \left( \nu^-,\nu^+,1 + \bar{\beta} ; x^+ \right)
\mbox{ , } F^-(r) = {_2F_1} \left( \mu^-,\mu^+,1 - \bar{\beta} ; x^+ \right) \label{Fp_Fm_PTG}.
\end{eqnarray}
Their arguments and employed parameters are:
\begin{eqnarray}
&&\bar{\beta} = -\frac{ik}{\Lambda^2 s} \mbox{ , } \nu^\pm = \frac{1}{2} \left( \ell + \frac{3}{2} + \bar{\beta} \pm \bar{\nu} \right)
\mbox{ , }\mu^\pm = \frac{1}{2} \left( \ell + \frac{3}{2} - \bar{\beta} \pm \bar{\nu} \right) , \\
&&\bar{\nu} = \sqrt{\left( \nu + \frac{1}{2} \right)^2 - \bar{\beta}^2 \left( \Lambda^2 (1-a) - 1 \right)} , \\
&&A^+ = \frac{\Gamma(\ell + \frac{3}{2}) \Gamma(- \bar{\beta})}{\Gamma(\mu^+) \Gamma(\mu^-)}
\mbox{ , } A^- = \frac{\Gamma(\ell + \frac{3}{2}) \Gamma(\bar{\beta})}{\Gamma(\nu^+) \Gamma(\nu^-)} , \\
&&x^+ = \frac{1-y^2}{1-y^2 + \Lambda^2 y^2} \mbox{ , } x^- = \frac{\Lambda^2 y^2}{1-y^2 + \Lambda^2 y^2}
\end{eqnarray}
where the $r$ dependence is directly visible in the $x^+$ and $x^-$ arguments via the $y$ parameter of Eqs.(\ref{y_equation_1},\ref{y_equation_2}).
From previous expressions, one can check that Eq.(\ref{PTG_wf2}) is obtained by operating on Eq.(\ref{PTG_wf1})
with the linear transformation formula of Eq.(\ref{T_mzp1}).
The alternative provided by Eqs.(\ref{PTG_wf1},\ref{PTG_wf2}) for the evaluation of $\varphi(r)$ is necessary as
Eq.(\ref{PTG_wf1}) and Eq.(\ref{PTG_wf2}) are respectively unstable for large and small $r$.
Eq.(\ref{PTG_wf2}) is also the usual expression of the wave function in terms of outgoing (``$+$'' terms)
and incoming part (``$-$'' terms), so that $A^- = 0$ for S-matrix pole states.
Occurrences of integer $b-a$ or $c-a-b$ values in $_2F_1(a,b,c;z)$ expressions of Eqs.(\ref{F_PTG},\ref{Fp_Fm_PTG})
can be verified to be possible for $\varphi(r)$ scattering (see Eqs.(\ref{PTG_wf1},\ref{PTG_wf2})) and for a discrete set of linear momenta $k \in \C$.

Scattering states are normalized according to the Dirac delta normalization. The $\mathcal{N}$ normalization reads:
\begin{eqnarray}
\mathcal{N} &=& \sqrt{\frac{2 \Lambda^2 s \bar{\beta} (\ell + \frac{3}{2} + \bar{\beta} + 2n) \Gamma(\ell + \frac{3}{2} + \bar{\beta} + n)
\Gamma(\ell + \frac{3}{2} + n)}
{(\ell + \frac{3}{2} + \bar{\beta} \Lambda^2 (1-a) + 2n) \Gamma(n+1) \Gamma(\bar{\beta} + n + 1) \Gamma(\ell + \frac{3}{2})^2}}
\mbox{ (poles)} \\
&=& \sqrt{\frac{\Gamma(\nu^+) \Gamma(\nu^-) \Gamma(\mu^+) \Gamma(\mu^-)}
{2 \pi \; \Gamma(\bar{\beta}) \Gamma(- \bar{\beta}) \Gamma(\ell + \frac{3}{2})^2}}
\mbox{ (scattering states) .}
\end{eqnarray}

Asymptotic expressions for $\varphi(r)$ for $r \rightarrow 0$ and $r \rightarrow +\infty$ are obtained from
Eqs.(\ref{y_r_small},\ref{y_r_large_1},\ref{y_r_large_2}) and Eqs.(\ref{PTG_wf1},\ref{PTG_wf2}):
\begin{eqnarray}
&&\varphi(r) \sim \mathcal{N} \; \sqrt{\Lambda(1-a)} (\Lambda s r)^{\ell+1} = C_0 r^{\ell+1} \mbox{ , } r \rightarrow 0 , \label{C0} \\
&&\varphi(r) \sim \mathcal{N} \; (A^+ e^{ik(r-r_1)} + A^- e^{-ik(r-r_1)}) = C^+ e^{ikr} + C^{-} e^{-ikr} \mbox{ , } r \rightarrow +\infty \label{Cplus_Cminus}
\end{eqnarray}
with:
\begin{eqnarray}
r_1 &=& \frac{1}{\Lambda^2 s} \left( \sqrt{\Lambda^2 - 1} \mbox{ } \arctan( \sqrt{\Lambda^2 - 1}) - \log \frac{\Lambda}{2} \right) \mbox{ , } \Lambda > 1  \\
    &=& -\frac{1}{\Lambda^2 s} \left( \sqrt{1 - \Lambda^2} \mbox{ arctanh}( \sqrt{1 - \Lambda^2}) + \log \frac{\Lambda}{2} \right) \mbox{ , } \Lambda \leq 1 .
\end{eqnarray}

The code also provides the first and second derivatives of the PTG wave functions, expressible with ${_2F_1}$ functions as well.

\section{The program \textit{hyp\_2F1}} \label{hyp_2F1_prog}

The code \textit{hyp\_2F1} is written in two independent versions, standard C++ and Fortran 90. 
It uses only standard libraries and is thus portable on many machines.

The C++ code is separated into three different files: \textit{complex\_functions.H}, \textit{hyp\_2F1.cpp} and \textit{hyp\_2F1\_example.cpp}
in order to ensure compatibility with the code of Ref.\cite{cwfcomplex}. Indeed, to use this code and Ref.\cite{cwfcomplex} simultaneously,
one has to use the present \textit{complex\_functions.H} instead of the file provided in Ref.\cite{cwfcomplex}.
The Fortran 90 code for ${_2F_1}$ is situated in \textit{hyp\_2F1.f90} and in \textit{hyp\_2F1\_example.f90}.

\textit{complex\_functions.H} contains elementary complex functions which are not in the standard library,
and routines calculating constants specific to the Coulomb wave functions for compatibility with Ref.\cite{cwfcomplex}.
The only differences with Ref.\cite{cwfcomplex} are situated in \textit{log\_Gamma}, calculated with full numerical precision instead of $10^{-10}$ precision,
and in the added function \textit{Gamma\_inv}, which calculate precisely $\displaystyle \frac{1}{\Gamma}(z)$.
Only the functions of the \textit{complex\_functions.H} file needed for the ${_2F_1}$ algorithm appear in \textit{hyp\_2F1.f90}, 
to which elementary functions which are present in C++ standard libraries but absent in Fortran 90, are added.

In \textit{hyp\_2F1.cpp} and \textit{hyp\_2F1.f90}, routines calculating the ${_2F_1}$ function consist of:
\begin{itemize}
\item \textit{Gamma\_ratio\_diff\_small\_eps}: calculate $H_{\epsilon}(z)$ for $0 \leq |\epsilon|_{\infty} \leq 0.1$ (see Eq.(\ref{H_eps_z})).
\item \textit{Gamma\_inv\_diff\_eps}: calculate $G_{\epsilon}(z)$ (see Eq.(\ref{G_eps_z})).
\item \textit{A\_sum\_init}: calculate $\displaystyle \frac{1}{\epsilon \; \Gamma(1-m-\epsilon)}$ of the $\alpha_0$ term of $A(z)$
(see Eqs.(\ref{a0_mzp1},\ref{a0_z_inv})).
\item \textit{log\_A\_sum\_init}: calculate directly the logarithm of \textit{A\_sum\_init} in case of overflow of the latter.
\item \textit{B\_sum\_init\_PS\_one}: calculate $\displaystyle \frac{\beta_0}{(1-z)^m}$ in the $B(z)$ sum for the $1-z$ linear transformation formula 
(see Eq.(\ref{b0_mzp1})).
\item \textit{B\_sum\_init\_PS\_infinity}: calculate $\displaystyle \frac{\beta_0}{z^{-m}}$ in the $B(z)$ sum 
for the $\displaystyle \frac{1}{z}$ linear transformation formula (see Eq.(\ref{b0_z_inv})).
\item \textit{cv\_poly\_der\_tab\_calc}: calculate the coefficients of the derivative of the $P(x)$ polynomial of Sec.(\ref{PS_cv_test}).
\item \textit{cv\_poly\_der\_calc}: evaluate the derivative of the $P(x)$ polynomial of Sec.(\ref{PS_cv_test}) for a given $x$ argument.
\item \textit{min\_n\_calc}: calculate the integer $n_c$ of Sec.(\ref{PS_cv_test}) ($\Delta > 0$) or return zero ($\Delta \leq 0$)
(see Sec.(\ref{PS_cv_test}) for method and notations).
\item \textit{hyp\_PS\_zero}: calculate $F(z)$ for $|z| < 1$ with Eqs.(\ref{hyp_PS_zero_2},\ref{hyp_PS_zero_term}).
\item \textit{hyp\_PS\_one}: calculate $F(z)$ for $|1-z| < 1$ with the method of Sec.(\ref{mzp1_transformation}).
\item \textit{hyp\_PS\_infinity}: calculate $F(z)$ for $|z| > 1$ with the method of Sec.(\ref{z_inv_transformation}).
\item \textit{hyp\_PS\_complex\_plane\_rest}: calculate $F(z)$ with a Taylor series expansion (see Eq.(\ref{Taylor_rest})).
\item \textit{hyp\_2F1}: calculate $F(z)$ for arbitrary $z$ with the method of Sec.(\ref{hyp_2F1_algo}).
\item \textit{test\_2F1}: test of $F(z)$ accuracy from Eqs.(\ref{2F1_test},\ref{2F1_test_one}).
\end{itemize}

The routines meant to be used directly are \textit{hyp\_2F1} and \textit{test\_2F1}.
\textit{hyp\_2F1} demands as arguments the complex values of $a$, $b$, $c$, and $z$ in this order, and return the calculated ${_2F_1}$ function.
To use \textit{test\_2F1}, the tested complex value $F(z)$ coming from \textit{hyp\_2F1} has to follow $a$, $b$, $c$ and $z$.
A real number measuring its accuracy is returned, determined via Eqs.(\ref{2F1_test},\ref{2F1_test_one}).
$F'(z)$ and $F''(z)$ are therein calculated directly with \textit{hyp\_2F1}.

It is possible to use \textit{Gamma\_inv\_diff\_eps}, 
\textit{hyp\_PS\_zero}, \textit{hyp\_PS\_one}, and \textit{hyp\_PS\_infinity} by themselves, but, for the three last functions, one has to pay
attention to the modulus of their $z$-parameter. One must have $|z| < 1$ and $|z| > 1$ respectively for \textit{hyp\_PS\_zero} and \textit{hyp\_PS\_infinity}.
\textit{hyp\_PS\_one} takes $1-z$ as argument instead of $z$, so that the very close vicinity of $z = 1$ can be handled precisely. $|1-z| < 1$ has
to be verified as well. Moreover, it is necessary to ensure that $\Re(c-a-b) \geq 0$ in \textit{hyp\_PS\_one} and $\Re(b-a) \geq 0$ in \textit{hyp\_PS\_infinity}.
All other functions are not supposed to be handled by the user.

Test examples are given in \textit{hyp\_2F1\_example.cpp} and \textit{hyp\_2F1\_example.f90}. It deals with a simple test of the $_2F_1$ function,
where the parameters $a$, $b$, $c$, and $z$ are given as inputs and the $_2F_1$ function evaluated value, as well the accuracy test of \textit{test\_2F1}, 
are provided as outputs.

\section{The program \textit{PTG\_wf}}

The PTG potential and wave functions are programmed differently according to the C++ or Fortran 90 codes.
In the C++ program, a class is defined for the potential, while a module is used for it in the Fortran 90 code. They are, however, very close
so that the C++ routines will be delineated first, and one will just point out differences between both versions afterward.

\subsection{Routines of the C++ program}
All routines related to the PTG potential are situated in \textit{PTG\_wf.cpp} and \textit{PTG\_wf\_example.cpp}

The class defining the PTG potential \textit{PTG\_class} is built from a constructor demanding the following arguments in that order:
\begin{itemize}
\item \textit{kin\_fact}: kinetic factor equal to $\displaystyle \frac{2m_0}{\hbar^2}$, where $m_0$ is the rest mass of the particle.
$\displaystyle \frac{\hbar^2}{2m_0}$ has the dimension of an energy times the square of a length.
\item \textit{l}: integer angular momentum of the PTG potential.
\item \textit{Lambda,s,nu,a}: parameters of the PTG potential defined in Sec.(\ref{PTG_eq_section}).
\end{itemize}
This class possesses the main functions pertaining to the PTG potential formulas of Sec.(\ref{PTG_eq_section}):
\begin{itemize}
\item \textit{Lambda2\_sr\_function}: calculate the function $\Lambda^2 s r(y)$ given by Eqs.(\ref{y_equation_1},\ref{y_equation_2}).
                                      The $y$ value must be given as argument.
\item \textit{y\_search}: calculate the $y(r)$ function of Sec.(\ref{y_calculation}) using Newton method and bisection.
A value of one is returned if $y_d > 0.99$ (see Sec.(\ref{y_calculation})).
\item \textit{exp\_minus\_2x\_iterative}: calculate $e^{-2x}$, with $x=\mbox{arctanh}(y)$ when $y \sim 1$ (see Sec.(\ref{y_calculation})).
This function and the precedent take $\Lambda^2 s r$ as argument.
\item \textit{effective\_mass}, \textit{effective\_mass\_der}: calculate the dimensionless effective mass $\mu(r)$ of the PTG potential (see Eq.(\ref{mu_r})).
$r$ is given as argument. Same for its derivative $\mu'(r)$.
\item \textit{V\_cl\_calc}: calculate the $\ell$-dependent $V_{\ell}(r)$ potential part 
with Eqs.(\ref{Vl_r},\ref{stable_Vl_r},\ref{PS_for_Vl},\ref{Vl_0}) in energy units.
\item \textit{operator()}: calculate the PTG potential $V_{PTG}(r)$ of Eq.(\ref{V_PTG}) in energy units.
\end{itemize}
$r$ always has the dimension of a length. Units can be chosen arbitrarily by the user and are invisible in the code.
All arguments and returned values are double-precision real. Member functions \textit{Lambda2\_sr\_function} and \textit{V\_cl\_calc} are private
as not intended to be used directly by the user.

PTG wave functions and the associated test are calculated with non-member routines:
\begin{itemize}
\item \textit{PTG\_pole}: calculate the wave function, derivative and second derivative of a particle S-matrix pole state of principal quantum number $n$
of a given PTG potential (see Eqs.(\ref{PTG_wf1},\ref{PTG_wf2})). It can be bound, resonant, or antibound.
Arguments are respectively: a reference on the \textit{PTG\_class}, the integer $n$, the number of points of the wave function,
a table of real $r$ radii, two complex numbers which will contain the $C_0$ and $C^+$ values of
Eqs.(\ref{C0},\ref{Cplus_Cminus}), and three tables of complex numbers where wave function and derivatives are stored.
\item \textit{PTG\_scat}: Same for a scattering state of complex linear momentum $k$. Arguments are the same except for $n$ which is replaced by $k$,
and a complex number is added between $C^+$ and wave function tables to store the value of $C^-$ from Eq.(\ref{Cplus_Cminus}).
\item \textit{PTG\_test\_calc}: calculate a test of the PTG wave function accuracy from its Schr{\"o}dinger equation (see Eq.(\ref{PTG_equation})).
Its arguments are: a reference on the \textit{PTG\_class}, the complex linear momentum $k$, a radius $r$ and three complex numbers
wielding the wave function, derivative and second derivative at $r$ radius.
\end{itemize}
Added to that, the complex function \textit{k\_PTG\_calc} returns the complex linear momentum of a S-matrix pole state of principal quantum number $n$,
taking as arguments the integers \textit{n} and \textit{l} and the real numbers \textit{Lambda}, \textit{s}, \textit{nu}, \textit{a}, and \textit{kin\_fact}
defining the PTG potential. Its unit is the inverse of a length.

\textit{PTG\_wf\_example.cpp} provides a test example of PTG wave functions. All parameters demanded by the class \textit{PTG\_class}
are input data, as well as the radial interval where PTG wave functions will be evaluated. The number of points considered in this interval are uniformly
distributed. Two states, a S-matrix pole and scattering state are calculated, so that the principal quantum number of the former and the linear momentum
of the latter are demanded as inputs. The output consists of the evaluated PTG wave functions, first and second derivative, as well as
in the PTG potential, effective mass and derivative at the same point $r$, and the accuracy test of \textit{PTG\_test\_calc}. The input file
has been configured so that the two possible formulas for calculating the PTG wave function $\varphi(r)$ in Eqs.(\ref{PTG_wf1},\ref{PTG_wf2}) are utilized.

\subsection{Fortran 90 program}
The Fortran 90 program routine names are the same, except for the \textit{operator()} function of the class \textit{PTG\_class}
which is renamed as \textit{V\_PTG}. They can be found in \textit{PTG\_wf.f90}.
Parameters of the PTG potential are initialized with the subroutine \textit{PTG\_PARAMETERS}, taking
the same arguments as the constructor of the class \textit{PTG\_class}. Arguments of other functions remain the same, except for the \textit{PTG\_class}
member reference of C++ routines, absent in the Fortran 90 code.

\section{Recommendations} \label{recommendations}
When utilizing \textit{hyp\_2F1}, the main issue is the size of $a$, $b$, and $c$ moduli.
While the code provides virtually full numerical precision for all values of $|a|$, $|b|$, and $|c|$ smaller or of the order of one,
precision starts to deteriorate for larger moduli (see Tabs.(\ref{table_test_1},\ref{table_test_2},\ref{table_test_3})).
Spurious effects are the most important when the imaginary parts of parameters $a$, $b$, and $c$ increase.
However, the average error is smaller than the maximal error by several orders of magnitude for moderate and large $|a|$, $|b|$, and $|c|$
(see Tabs.(\ref{table_test_1},\ref{table_test_2},\ref{table_test_3})).
This indicates that the instabilities are very localized in the $a$, $b$, $c$,  and $z$ four-dimensional complex space. Hence,
calculations are usually meaningful for moderate $|a|$, $|b|$, and $|c|$.
Testing functions is therein necessary to ensure the validity of the calculation, rendering, however, the time of calculation three times slower due
to ${_2F_1}$ calculations overhead. If accuracy is poor, it is possible to use directly \textit{hyp\_PS\_zero},
\textit{hyp\_PS\_one}, and \textit{hyp\_PS\_infinity} along with transformations of Sec.(\ref{linear_transf_section}) instead of \textit{hyp\_2F1},
as mentioned in Sec.(\ref{hyp_2F1_prog}), as the method automatically chosen by the code to evaluate $_2F_1$ may not be the most robust
for some particular instances of $a$, $b$, $c$, and $z$. The polynomial case is almost devoid of numerical inaccuracy for polynomial degrees
smaller than ten (see Tab.(\ref{table_test_poly})).

If one calculates \textit{hyp\_2F1} functions for integer-spaced $a$, $b$, and $c$, the recurrence relations of Ref.\cite{Abramowitz_2F1}
could be helpful (see also Ref.\cite{Vidunas} for a theoretical study of these relations).
However, they become unstable if the calculated ${_2F_1}$ functions are minimal solutions of Eq.(\ref{hyp_eq}) \cite{Temme_paper}.
Hence, the calculation of ${_2F_1}(a+n,b+m,c+p;z)$ values, where $(n,m,p) \in \Z$,
theoretically reachable from ${_2F_1}(a \pm 1,b,c;z)$, ${_2F_1}(a,b \pm 1,c;z)$ and ${_2F_1}(a,b,c \pm 1;z)$
evaluations and recurrence relations \cite{Abramowitz_2F1}, would become too difficult to code. If the calculated ${_2F_1}$ function is minimal,
backward recurrence with Miller's algorithm has to be used, while forward recurrence must be employed in the opposite case \cite{Temme_paper}.
As all three parameters $a$, $b$, and $c$ must be recurred, a general algorithm allowing the evaluation of ${_2F_1}(a+n,b+m,c+p;z)$
would have to consider all possible forward and backward recurrences which can occur for arbitrary complex $a$, $b$, and $c$, which would demand
to consider too many particular cases.
Moreover, they cannot solve the instability problem appearing for large $|\Im[a]|$, $|\Im[b]|$, and $|\Im[c]|$.
They, may, however, prove useful for some particular problems.

The situation concerning PTG wave functions is much better. Firstly, PTG pole states of principal quantum number $n$ only involve ${_2F_1}$ polynomials
of degree $n$ (Jacobi polynomials) \cite{PTG_pot}. $n$ rarely exceeds ten, so that the polynomial case of ${_2F_1}$ is therein sufficient.
In case of the opposite, the Jacobi polynomials can be evaluated with stable recurrence relations. However, this situation is considered
too particular to be included in our general code. Secondly, ${_2F_1}$ parameters of PTG scattering states will
usually be of a small modulus, so that the aforementioned instabilities will be absent in practice.

\section{Conclusion}
The direct evaluation of the ${_2F_1}$ function with power series and all its parameters and argument complex has been attempted for the first time.
The proposed method, relying on linear transformation theory and analytical properties of the Lanczos approximation of the Gamma function,
has proved to be very efficient and allows to access all $z$ values of the complex plane, the vicinity of negative integers $b-a$ and $c-a-b$
being rigorously treated. Its main drawback is its propensity to become unstable in general when $|a|$, $|b|$, and $|c|$ increase. Such instabilities
are, however, very localized, so that the code can still be used in most cases, as long as the accuracy of the functions has been verified to
be satisfactory. Despite this obvious disadvantage, this code will surely allow to render possible calculations which before demanded the implementation of
complicated algorithms for each encountered particular situation.

The evaluation of PTG wave functions has been provided as a physical application,
as ${_2F_1}$ functions appear directly in their analytical formulas and because its implementation demands care due to its implicitly defined arguments.
PTG wave functions have the advantage of being both analytical and bear bound, resonant and scattering states
due to the finite range of the PTG potential. The published code permits a fast and stable evaluation of PTG wave functions, which will
probably be of interest for studies in the domain of loosely bound and particle-emitting quantum systems.

\section*{Acknowledgments}
Discussions with K.~Matsuyanagi are gratefully acknowledged.
The authors acknowledge Japan Society for the Promotion of Science for 
awarding The JSPS Postdoctoral Fellowship for Foreign Researchers and The Invitation Fellowship for Research in Japan (Long-term)
This work was supported in part by the U.S.~Department of Energy under Contracts Nos.~DE-FG02-96ER40963 (University of Tennessee), 
DE-AC05-00OR22725 with UT-Battelle, LLC (Oak Ridge National Laboratory), and DE-FG05-87ER40361 (Joint Institute for Heavy Ion Research), 
DE-FC02-07ER41457 (University of Washington).

\newpage
\begin{table}[htbp]
\begin{center}
\caption{Relative accuracy of the $_2F_1$ function obtained via Eqs.(\ref{2F1_test},\ref{2F1_test_one})
for $\displaystyle z = r e^{i \frac{\pi}{3}}$, with $r = 0.99$.
The methods of Eqs.(\ref{Taylor_rest},\ref{Taylor_rest_iter_coeff}), denoted as ``ours'',
and the scheme of Ref.\cite{Buhring}, denoted as ``B{\"u}hring'', are compared.
$|\Re(a,b,c)|$ represents either $|\Re(a)|$, $|\Re(b)|$ or $|\Re(c)|$, so that
0-1 for $|\Re(a,b,c)|$ means that for $x \in \{a,b,c\}$,  $0 \leq |\Re(x)| \leq 1$.
For each line of the table, $30,000$ $(a,b,c)$ sets of parameters
are considered and are chosen randomly so that $|\Im(a,b,c)| \leq 1$,
with $|\Im(a,b,c)|$ defined similarly to $|\Re(a,b,c)|$.
$T_{max}$ is the maximal value obtained with Eqs.(\ref{2F1_test},\ref{2F1_test_one})
and $T_{av}$ is the average of all values given by Eqs.(\ref{2F1_test},\ref{2F1_test_one}).}
\vspace*{1cm}
\begin{tabular}{|c|c|c|c|c|} \hline
$|\Re(a,b,c)|$ & $T_{max}$ (ours) & $T_{max}$ (B{\"u}hring) & $T_{av}$ (ours) & $T_{av}$ (B{\"u}hring) \\ \hline
0-1   & $ 9.2 \; 10^{-15} $ & $ 1.4 \; 10^{-11} $ & $ 6.1 \; 10^{-16} $ & $ 4.7 \; 10^{-15} $ \\
1-2   & $ 9.7 \; 10^{-15} $ & $ 1.0 \; 10^{-10} $ & $ 1.1 \; 10^{-15} $ & $ 3.5 \; 10^{-14} $ \\
2-5   & $ 5.2 \; 10^{-14} $ & $ 5.1 \; 10^{-10} $ & $ 1.6 \; 10^{-15} $ & $ 1.6 \; 10^{-13} $  \\
5-10  & $ 9.5 \; 10^{-13} $ & $ 3.6 \; 10^{-6} $  & $ 6.7 \; 10^{-15} $ & $ 2.4 \; 10^{-9} $ \\
10-15 & $ 3.2 \; 10^{-11} $ & $ 1.1 \; 10^{-2} $  & $ 1.7 \; 10^{-13} $ & $ 2.9 \; 10^{-5} $ \\ \hline
\end{tabular}
\label{table_Buhring_test_1}
\end{center}
\end{table}

\begin{table}[htbp]
\begin{center}
\caption{Same as in Tab.(\ref{table_Buhring_test_1}), but with $r = 1.01$.}
\vspace*{1cm}
\begin{tabular}{|c|c|c|c|c|} \hline
$|\Re(a,b,c)|$ & $T_{max}$ (ours) & $T_{max}$ (B{\"u}hring) & $T_{av}$ (ours) & $T_{av}$ (B{\"u}hring) \\ \hline
0-1   & $ 9.0 \; 10^{-13} $ & $ 9.0 \; 10^{-12} $ & $ 3.0 \; 10^{-15} $ & $ 4.1 \; 10^{-15} $ \\
1-2   & $ 4.0 \; 10^{-11} $ & $ 9.2 \; 10^{-11} $ & $ 2.2 \; 10^{-14} $ & $ 3.6 \; 10^{-14} $ \\
2-5   & $ 9.6 \; 10^{-11} $ & $ 3.7 \; 10^{-10} $ & $ 6.8 \; 10^{-14} $ & $ 1.5 \; 10^{-13} $  \\
5-10  & $ 4.2 \; 10^{-9} $  & $ 3.0 \; 10^{-6} $  & $ 1.8 \; 10^{-12} $ & $ 1.5 \; 10^{-9} $ \\
10-15 & $ 8.5 \; 10^{-7} $  & $ 6.9 \; 10^{-2} $  & $ 2.6 \; 10^{-10} $ & $ 1.6 \; 10^{-5} $ \\ \hline
\end{tabular}
\label{table_Buhring_test_2}
\end{center}
\end{table}

\begin{table}[htbp]
\begin{center}
\caption{Relative accuracy of the $_2F_1$ function obtained via Eqs.(\ref{2F1_test},\ref{2F1_test_one}).
For each line of the table, $10^6$ $(a,b,c,z)$ sets of parameters
are considered and are chosen randomly so that $|z|_{\infty} \leq 3$ and $|\Im(a,b,c)| \leq 1$.
$|\Re(a,b,c)|$, $|\Im(a,b,c)|$, $T_{max}$ and $T_{av}$ are defined similarly as in Tab.(\ref{table_Buhring_test_1}).}
\vspace*{1cm}
\begin{tabular}{|c|c|c|} \hline
$|\Re(a,b,c)|$ & $T_{max}$ & $T_{av}$ \\ \hline
0-1   & $ 1.0 \; 10^{-12} $ & $ 3.0 \; 10^{-16} $ \\
1-2   & $ 4.1 \; 10^{-11} $ & $ 2.0 \; 10^{-15} $ \\
2-5   & $ 3.1 \; 10^{-10} $ & $ 2.4 \; 10^{-14} $ \\
5-10  & $ 5.0 \; 10^{-5}  $ & $ 8.8 \; 10^{-11} $ \\
10-15 & $ 2.6             $ & $ 1.2 \; 10^{-5} $ \\ \hline
\end{tabular}
\label{table_test_1}
\end{center}
\end{table}

\begin{table}[htbp]
\begin{center}
\caption{Same as Tab.(\ref{table_test_1}) except that the condition $1 \leq |\Im(a,b,c)| \leq 2$ is verified.}
\vspace*{1cm}
\begin{tabular}{|c|c|c|} \hline
$|\Re(a,b,c)|$ & $T_{max}$ & $T_{av}$ \\ \hline
0-1   & $ 1.3 \; 10^{-10} $ & $ 7.7 \; 10^{-15} $ \\
1-2   & $ 9.2 \; 10^{-9}  $ & $ 1.7 \; 10^{-13} $ \\
2-5   & $ 3.3 \; 10^{-5}  $ & $ 6.6 \; 10^{-10} $ \\
5-10  & $ 8.7 \; 10^{-3}  $ & $ 2.2 \; 10^{-8}  $ \\
10-15 & $ 5.8             $ & $ 4.4 \; 10^{-5}  $ \\ \hline
\end{tabular}
\label{table_test_2}
\end{center}
\end{table}

\begin{table}[htbp]
\begin{center}
\caption{Same as Tab.(\ref{table_test_1}) except that the condition $2 \leq |\Im(a,b,c)| \leq 5$ is verified.}
\vspace*{1cm}
\begin{tabular}{|c|c|c|} \hline
$|\Re(a,b,c)|$ & $T_{max}$ & $T_{av}$ \\ \hline
0-1   & $ 7.0 \; 10^{-2}  $ & $ 3.4 \; 10^{-7} $ \\
1-2   & $ 1.5 \; 10^{-1}  $ & $ 2.7 \; 10^{-7} $ \\
2-5   & $ 7.3             $ & $ 1.5 \; 10^{-4} $ \\ \hline
\end{tabular}
\label{table_test_3}
\end{center}
\end{table}

\begin{table}[htbp]
\begin{center}
\caption{Relative accuracy of $_2F_1$ polynomials obtained by way of Eqs.(\ref{2F1_test},\ref{2F1_test_one}).
         $_2F_1$ functions are polynomials of degree running from zero to ten,
         so that $a$ is fixed to minus the degree. Parametrization is the same as in Tab.(\ref{table_test_1}),
         except that $\Re(b,c)$ and $\Im(b,c)$ are used instead of $\Re(a,b,c)$ and $\Im(a,b,c)$. 
         One has here $|\Im(b,c)| \leq 10$.}
\vspace*{1cm}
\begin{tabular}{|c|c|c|} \hline
$|\Re(b,c)|$ & $T_{max}$ & $T_{av}$ \\ \hline
0-1   & $ 5.1 \; 10^{-8} $ & $ 1.2 \; 10^{-12} $ \\
1-2   & $ 8.1 \; 10^{-7} $ & $ 2.3 \; 10^{-12} $ \\
2-5   & $ 4.1 \; 10^{-8} $ & $ 2.3 \; 10^{-12} $ \\
5-10  & $ 3.7 \; 10^{-7} $ & $ 8.9 \; 10^{-12} $ \\
10-15 & $ 3.8 \; 10^{-6} $ & $ 3.9 \; 10^{-11} $ \\ \hline
\end{tabular}
\label{table_test_poly}
\end{center}
\end{table}


\begin{thebibliography}{10}

\bibitem{cwfcomplex} N.~Michel, Comp. Phys. Comm., 176 (2007) 232.
\bibitem{Alder_Bohr} K.~Alder, A.~Bohr, T.~Huus, B.~Mottelson, and A.~Winther, Rev. Mod. Phys., 28 (1956) 432.
\bibitem{coulex_book} L.C.~Biedenharn and P.J.~Brussaard, ``Coulomb Excitation'', Clarendon, Oxford (1965).
\bibitem{coulex_paper} C.J.~Mullin and E.~Guth, Phys. Rev., 82 (1951) 141.
\bibitem{van_ingen} H.~van Haeringen and R.~van Wageningen, J. Math. Phys., 16 (1975) 1441.
\bibitem{Datta} S.~Datta, J. Phys. B:At. Mol. Phys., 18 (1985) 853.
\bibitem{Kepler_classical} G.~Adkins and J.~Mc~Donnell, Phys. Rev. D, 75 (2007) 082001.
\bibitem{Kepler_relativistic} L.~Hostler, J. Math. Phys., 28 (1987) 2984.
\bibitem{linear_MHD} J.A.~Adam, J. Math. Phys., 30 (1989) 744.
\bibitem{bin_stars_grav_waves} V.~Pierro, I.M.~Pinto, and A.~Spallici, Mon. Not. R. Astron. Soc., 334 (2002) 855.
\bibitem{Eckart_pot} C.~Eckart, Phys. Rev., 35 (1930) 1303.
\bibitem{Rosen_Morse_pot} N.~Rosen and P.M.~Morse, Phys. Rev., 42 (1932) 210.
\bibitem{Hulthen_pot} L.~Hulth{\'e}n, Ark. Mat. Astron. Fys. A, 28 (1942) 5.
\bibitem{Manning_Rosen_pot1} M.F.~Mannning and N.~Rosen, Phys., Rev. 44 (1933) 953.
\bibitem{Manning_Rosen_pot2} S.H.~Dong and J.~Garcia-Ravelo, Phys. Scr., 75 (2007) 307.
\bibitem{Natanzon_pot1} G.A.~Natanzon, Vestn. Leningr. Univ. Fiz., Khim. 10 (1971) 22; Theor. Math. Phys., 38 (1979) 146.
\bibitem{Klein-Gordon} C.~Rojas and V.M.~Villalba, Phys. Rev. A, 71 (2005) 052101.
\bibitem{Dirac1} Guo Jian-You, Fang Xiang Zheng, and Xu Fu-Xin, Phys. Rev. A, 66 (2002) 062105.
\bibitem{Dirac2} Guo Jian-You, Meng Jie, and Xu Fu-Xin, Chin. Phys. Lett., 20 (2003) 602.
\bibitem{Dirac3} Y.~Sucu and N.~{\"U}nal, J. Math. Phys., 48 (2007) 052503.
\bibitem{Ginocchio_SUSY} F.~Cooper, J.N.~Ginocchio, and A.~Khare, Phys. Rev. D, 36 (1987) 2458.
\bibitem{Natanzon_SO21} P.~Cardero and S.~Salamo, J. Math. Phys., 35 (1994) 3301.
\bibitem{Natanzon_path_int} L.~Chetouani, L.~Guechi, A.~Lecheheb and T.~Hammann, J. Math. Phys., 34 (1993) 1257.
\bibitem{PTG_pot} J.N.~Ginocchio, Ann. Phys., 152 (1984) 203; 159 (1985) 467.
\bibitem{PTG_path_int} L.~Chetouani, L.~Guechi, A.~Lecheheb,  and T.~Hammann, Czech. J. Phys., 45 (1995) 699.
\bibitem{Karim} K.~Bennaceur, J.~Dobaczewski, and M.~P{\l}oszajczak, Phys. Rev. C, 60 (1999) 034308; Phys. Lett. B, 496 (2000) 154.
\bibitem{Mario1} M.V.~Stoitsov, S.~Dimitrova, S.~Pittel, P.~van~Isacker, and A.~Frank, Phys. Lett. B, 415 (1997) 1.
\bibitem{Mario2} S.~Pittel and M.V.~Stoitsov, J. Phys. G: Nucl. Part. Phys., 24 (1998) 1461.
\bibitem{Mario3} {M.V.~Stoitsov, N.~Michel, and K.~Matsuyanagi, nucl-th/0709.1006, submitted to Phys. Rev. C.}
\bibitem{Abramowitz_2F1} M.~Abramowitz, Chap.~15 ``Hypergeometric Functions'', Handbook of Mathematical Functions,
edited by M.~Abramowitz and I.A.~Stegun, National Bureau of Standards, Applied Mathematics Series - 55 (1972).
\bibitem{Forrey} R.C.~Forrey, J. Comp. Phys., 137 (1997) 79.
\bibitem{Temme_pres} ``Numerical Aspects of Special Functions'', N.M.~Temme, ``Numerics of Special Functions'', ICNAAM 2005, Rhodes, Greece, 16-20 September \\
                     http://www.cant.ua.ac.be/workshops/files/icnaam/temme.pdf
\bibitem{Buhring} W.~B{\"u}hring, J. SIAM Math. Anal., 18 (1987) 884.
\bibitem{Becken_Schmelcher} W.~Becken and P.~Schmelcher, J. Comp. Appl. Math., 126 (2000) 449.
\bibitem{NR} W.H.~Press, S.A.~Teukolsky, W.T.~Vetterling, and B.P.~Flannery, Numerical Recipes in C, Cambridge University Press 1988-1992.
\bibitem{Abramowitz_Gamma} M.~Abramowitz, Chap.~6 ``Gamma function and related functions'', Handbook of Mathematical Functions,
edited by M.~Abramowitz and I.A.~Stegun, National Bureau of Standards, Applied Mathematics Series - 55 (1972). 
\bibitem{Higham} N.J.~Higham, ``Accuracy and Stability of Numerical Algorithms'', SIAM, Philadelphia, 1996.
\bibitem{Temme_paper} A.~Gil, J.~Segura, and N.M.~Temme, Math. Comp., 76 (2007) 1449.
\bibitem{Vidunas} R.~Vidunas, J. Comp. Appl. Math., 153 (2003) 507.


\end{thebibliography}
\end{document}